\documentclass[twocolumn,amsmath,amssymb,superscriptaddress,nopacs,floats]{revtex4-1}

\usepackage[usenames,dvips]{color}
\usepackage{graphicx}
\usepackage{dcolumn}
\usepackage{bm}
\usepackage{pifont}
\usepackage{epstopdf}
\usepackage{xcolor}
\usepackage{mathtools}
\usepackage[bookmarks=false,colorlinks=true,urlcolor=blue,citecolor=blue,linkcolor=blue]{hyperref}
\usepackage{tikz}
\usepackage{enumitem}
\usepackage[normalem]{ulem}

\definecolor{mag_spiral}{RGB}{75,150,178}
\definecolor{inc_stripe}{RGB}{101,101,255}
\definecolor{collinear}{RGB}{172,99,56}
\definecolor{non_collinear}{RGB}{225,96,105}
\definecolor{spiral_w_stripe}{RGB}{35,107,99}
\definecolor{spiral_w_orth_stripe}{RGB}{255,255,255}
\definecolor{coplanar}{RGB}{184,78,140}
\definecolor{c4_noncoplanar}{RGB}{102,204,108}
\definecolor{c2_noncoplanar}{RGB}{220,201,108}

\definecolor{yess}{RGB}{0,100,0}

\newcommand{\bea}{\begin{eqnarray}}
\newcommand{\eea}{\end{eqnarray}}
\newcommand{\bt}{\textbf}

\newcommand{\phd}{\phantom{\dag}}
\newcommand{\ph}{\phantom{.}}

\newcommand{\noi}{\noindent}
\newcommand{\no}{\nonumber}

\newcommand{\oust}{{\color{red}\ding{55}}}
\newcommand{\yes}{{\color{yess}\ding{52}}}

\begin{document}
\def\v#1{{\bf #1}}

\title{Unravelling incommensurate magnetism and its emergence in\\ iron-based superconductors}

\author{Morten H. Christensen}
\affiliation{Niels Bohr Institute, University of Copenhagen, Juliane Maries Vej 30, 2100 Copenhagen, Denmark}
\author{Brian M. Andersen}
\affiliation{Niels Bohr Institute, University of Copenhagen, Juliane Maries Vej 30, 2100 Copenhagen, Denmark}
\author{Panagiotis Kotetes}
\affiliation{Niels Bohr Institute, University of Copenhagen, Juliane Maries Vej 30, 2100 Copenhagen, Denmark}
\affiliation{Center for Quantum Devices, Niels Bohr Institute, University of Copenhagen, 2100 Copenhagen, Denmark}

\begin{abstract}
We focus on a broad class of tetragonal itinerant systems sharing a tendency towards the spontaneous formation of incommensurate magnetism with ordering wavevectors $\bm{Q}_{1,2}=(\pi-\delta,0)/(0,\pi-\delta)$ or $\bm{Q}_{1,2}=(\pi,\delta)/(-\delta,\pi)$. Employing a Landau approach, we obtain the generic magnetic phase diagram and identify the lea\-ding instabi\-li\-ties near the paramagnetic-magnetic transition. Nine distinct magnetic phases exist that either preserve or violate the assumed C$_4$-symmetry of the paramagnetic phase. These are single- and double-$\bm{Q}$ phases consisting of magnetic stripes, helices and whirls, either in an individual or coexisting manner. These nine phases can be experimentally distinguished by polarized neutron scattering, or, for example, by combining measurements of the induced charge order and magnetoelectric coupling. Within two representative five-orbital models, suitable for BaFe$_2$As$_2$ and LaFeAsO, we find that the incommensurate magnetic phases discussed here are accessible in iron-based superconductors. Our investigation unveils a set of potential candidates for the unidentified C$_2$-symmetric magnetic phase that was recently observed in Ba$_{1-x}$Na$_x$Fe$_{2}$As$_{2}$. Among the phases stabilized we find a spin-whirl crystal, which is a textured magnetic C$_4$-symmetric phase. The possible experimental observation of textured magnetic orders in iron-based superconductor, opens new directions for realizing intrinsic to\-po\-lo\-gi\-cal superconductors.

\end{abstract}

\maketitle

\section{Introduction}

Magnetism constitutes one of the most ubiquitous phases in correlated matter, giving rise to a wide range of remarkable phenomena. Apart from the typical exam\-ples of (anti)ferromag\-ne\-tism, there exists an entire zoo of other magnetic phases, ranging from collinear to non-coplanar orders. The iron-based superconductors (FeSCs) are a prime example of this diversity. The commensurate magnetic stripe (MS) phase, with moments aligned antiferromagnetically along one Fe-Fe direction and ferromagnetically along the other, is prevalent in the undoped compounds. However, recent experiments on a number of hole-doped compounds have revealed the emergence of both collinear~\cite{allred16a} and coplanar~\cite{meier17} magnetic orders. 

Commensurate magnetism in the FeSCs can be described by two magnetic order pa\-ra\-me\-ters $\bm{M}_1$ and $\bm{M}_2$, with ordering vectors $\bm{Q}_1=(\pi,0)$ and $\bm{Q}_2=(0,\pi)$, related by fourfold (C$_4$) rotations~\cite{lorenzana08,eremin,brydon, giovannetti,gastiasoro15,wang15,kang15a,christensen15,christensen17}. In the event that the two order parameters compete, the single-$\bm{Q}$ MS phase is realized. On the other hand, if the two order parameters coexist the moments can align in an either parallel ($\bm{M}_1 \parallel \bm{M}_2$) or perpendicular ($\bm{M}_1 \perp \bm{M}_2$) fashion. In the case of collinear moments the resulting magnetic order is dubbed a charge-spin density wave (CSDW) phase, while the case of non-collinear moments is denoted a spin-vortex crystal (SVC) phase. These so-called double-$\bm{Q}$ phases can be challenging to distinguish experimentally, and this ge\-ne\-ral\-ly requires the use of local probe techniques. Signatures of double-$\bm{Q}$ magnetic structures were observed in Ba$_{1-x}$Na$_x$Fe$_{2}$As$_{2}$~\cite{avci14a,wasser15}, Ba$_{1-x}$K$_{x}$Fe$_{2}$As$_{2}$~\cite{hassinger,bohmer15a,allred15a,zheng16a,malletta,mallettb}, Sr$_{1-x}$Na$_{x}$Fe$_{2}$As$_{2}$~\cite{allred16a}, and in hole-doped CaKFe$_4$As$_4$~\cite{meier17}. In Sr$_{1-x}$Na$_x$Fe$_2$As$_2$ M{\"o}ssbauer spectroscopy captured the existence of a CSDW phase~\cite{allred16a}, while in CaKFe$_4$As$_4$ an SVC phase was observed~\cite{meier17}. Their disco\-ve\-ry has ge\-ne\-ra\-ted considerable attention and especially the observation of a CSDW phase highlights the itinerant nature of magnetism in these compounds.

Nevertheless, recent experiments suggest that magnetism in FeSCs may not only be li\-mi\-ted to the three phases mentioned above. In fact, the commensurate aspect needs to be revisited since direct evidence for incom\-mensurate (IC) magnetism has been provided by neutron scat\-te\-ring~\cite{pratt,luo,braden} on electron-doped FeSCs. In addition, recent thermal expansion measurements~\cite{wang16a} performed on hole-doped Ba$_{1-x}$Na$_x$Fe$_{2}$As$_{2}$ have revealed a rich mosaic of single- and double-$\bm{Q}$ magnetic phases, inclu\-ding a new C$_2$-symmetric phase of a currently unidentified nature. Importantly, the latter experiment found evidence for the presence of an inflection point in the magnetic transition temperature as a function of doping. This has been interpreted as the onset of IC magnetic order~\cite{Inflection,chubukov10}, although the presence of IC magnetism in this compound has not yet been confirmed by scattering experiments. We note that the existence of IC magnetic order is entirely expected within an itinerant scenario for magnetism in the FeSCs.

Similar to other high-$T_c$ superconductors~\cite{highTc,paglione10}, superconductivity in the FeSCs emerges via charge do\-ping of the parent compounds. This leads to the possi\-bi\-li\-ty of microscopic coexistence between superconduc\-ti\-vi\-ty and magnetism, which has indeed been observed in certain pnictogen compounds~\cite{johrendt11,avci11,klauss15,avci14a,wang16a,ni08a,nandi10a}. In the context of textured non-collinear and non-coplanar magnetic phases, such a coexistence is of particular interest. These types of orders tie together the orbital and spin degrees of freedom and induce an effective spin-orbit coupling (SOC)~\cite{Braunecker}. When the induced SOC breaks inversion symmetry, such magnetic orders can give rise to magnetoelectric phenomena, akin to those encountered in spin-skyrmion crystals~\cite{THE}, multiferroics~\cite{multiferroic}, semiconductors with non-negligible SOC~\cite{SpinGalvanic,SpinHallEffects}, and topological insulators~\cite{QiZhang}. More importantly, the simultaneous vio\-la\-tion of time-reversal symmetry and the generation of SOC renders the aforementioned magnetic textures sui\-table building blocks for rea\-li\-zing topological superconductors (TSCs)~\cite{Choy,Ivar,Flensberg,Stevan,KotetesClassi,Nakosai2013,Simon,Klinovaja, Franz,Pientka,Jian,ChristensenTSC,schecter16,Ojanen,Sedlmayr,Mendler,Chen,Loss}. This can be achieved either via pro\-xi\-mi\-ty of these magnetic textures to a conventional superconductor (SC), or intrinsically, by virtue of their microscopic coe\-xi\-sten\-ce with spin-singlet superconducti\-vi\-ty. An intrinsic TSC remains a long-sought-after but also elusive phase of matter, and ac\-com\-pli\-shing it via the above mechanism is challenging due to the anta\-go\-ni\-stic relation of magnetic order and spin-singlet superconductivity.

\begin{figure*}[t!]
\centering
\includegraphics[width=0.9\textwidth]{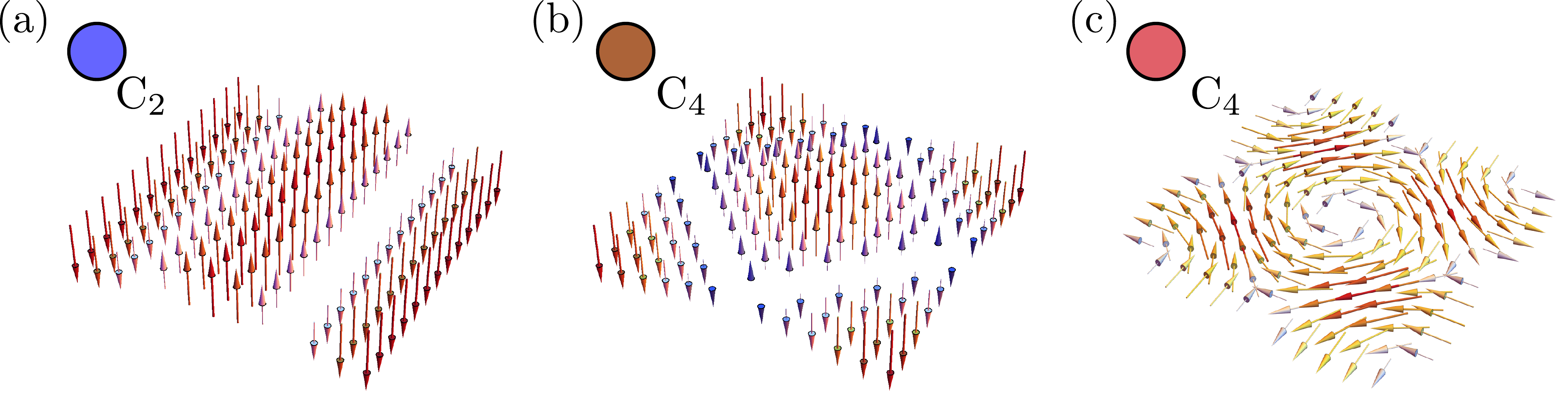}
\caption{\label{fig:IC_generalizations} Illustration of the IC generalization of the three well-known commensurate magnetic phases with ordering wavevectors $(0,\pi)$ and $(\pi,0)$. Here we restrict to a single magnetic unit cell and depict the spatial profile of the magnetization for: (a) the C$_2$-symmetric IC magnetic stripe (IC-MS) phase, (b) the C$_4$-symmetric IC charge-spin density wave (IC-CSDW) phase and (c) the C$_4$-symmetric IC spin-vortex crystal (IC-SVC) phase.}
\end{figure*}

\begin{figure*}[t!]
\centering
\includegraphics[width=0.9\textwidth]{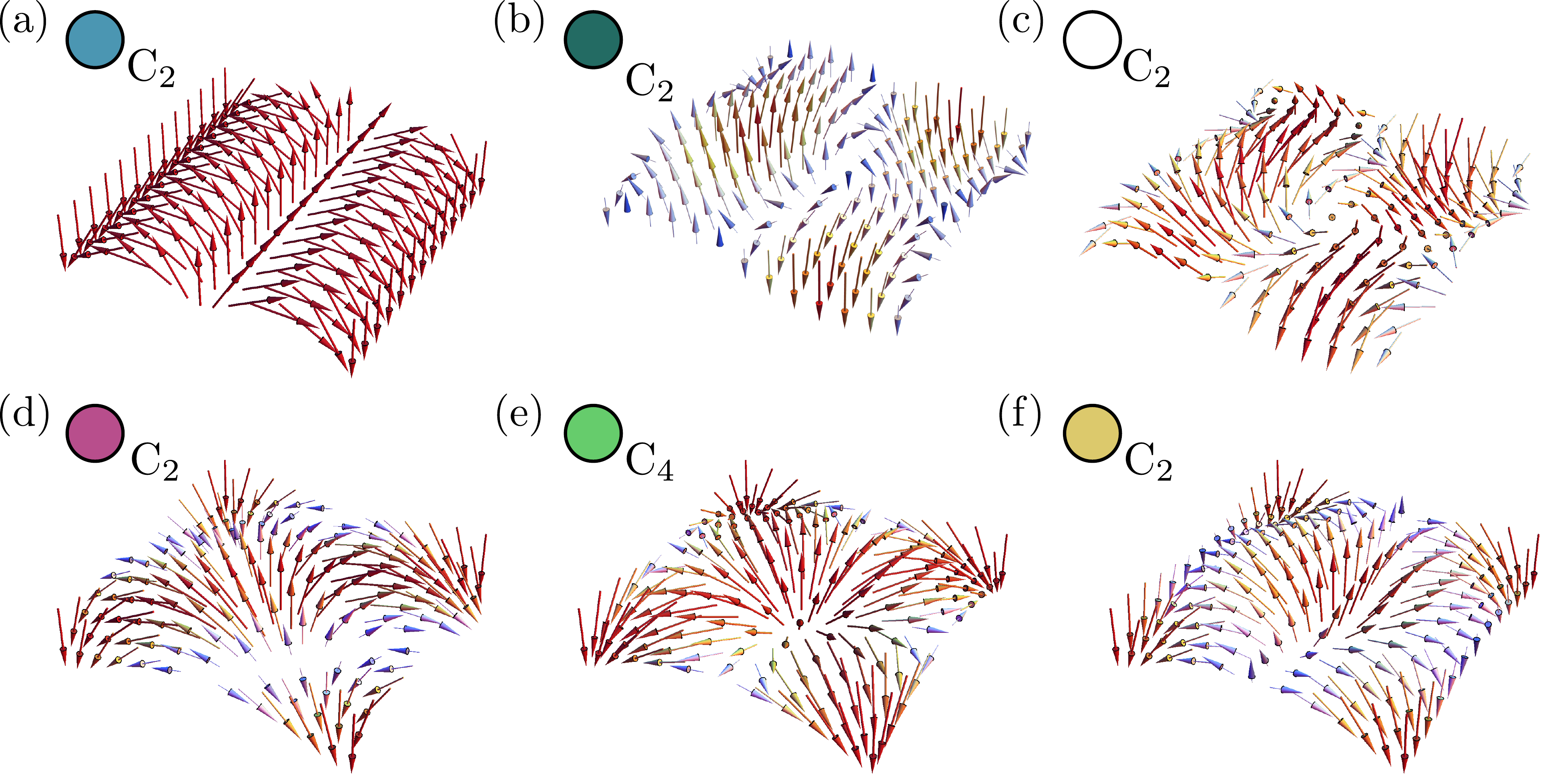}
\caption{\label{fig:IC_magnetic_phases} Novel mag\-netic phases appea\-ring with the rise of IC magnetism. The color scale signifies the magnitude of the magnetic moment. Here we restrict to a single magnetic unit cell. Contrary to the commensurate case, the incommensurability allows for non-coplanar magnetic textures. In (a)-(c) various magnetic helix (MH) order parameters are shown. In (b)/(c) the helix co\-exists with an in-/out-of-plane IC stripe (MS$||$MH/MS$\perp$MH). In (d) we present the C$_2$-symmetric double parallel magnetic helix (DPMH) phase and in (e)/(f) the non-coplanar spin-whirl crystal (SWC$_{4}$/SWC$_2$) phase with C$_4$/C$_2$ symmetry.}
\end{figure*}

Motivated by the recent experimental developments and the possible applications for TSCs, we explore the generic consequences of an IC magnetic ordering vector. Interestingly, we find that this leads to a number of magnetic phases distinct from the previously discussed stripe, collinear, and non-collinear orders. For this purpose, we first carry out a sy\-ste\-ma\-tic exploration of the accessible IC magnetic orders with wavevectors $\bm{Q}_{1,2}=(\pi-\delta,0)/(0,\pi-\delta)$ or $\bm{Q}_{1,2}=(\pi,\delta)/(-\delta,\pi)$. Within a generic Landau approach we identify nine possible IC magnetic ground states, and subsequently extract the phase diagram. The appearance of new phases demonstrates that the effect incommensurability is not limited to a simple generalization of the three known commensurate phases. Among the phases we find a single-$\bm{Q}$ magnetic helix phase along with two double-$\bm{Q}$ phases where an IC stripe coe\-xists with a magnetic helix. Additionally, we find a double-$\bm{Q}$ phase consisting of two parallel helices, and, finally, two double-$\bm{Q}$ phases consti\-tu\-ting, as coined here, the spin-whirl crystals. The fundamental properties of all the nine magnetic phases are summarized in Table~\ref{table:ICMagneticSignatures}, while the respective spatial profiles of the magnetization are depicted in Figs.~\ref{fig:IC_generalizations} and \ref{fig:IC_magnetic_phases}, and discussed in greater detail below.

\begin{table*}[t!]\caption{Characteristics and experimental signatures of the nine distinct incommensurate magnetic phases, which are extrema of Eq.~\eqref{eq:free_energy}. The wavevectors appearing in the column regarding the induced charge order have been separated into groups. For a given group the Bragg peaks have the same intensity. Note that the wavevectors in gray, depending on the parameters, may not lead to a Bragg peak in the spectrum. $\bm{{\cal B}}/\bm{{\cal E}}$ and ${\cal C}$ denote a Zeeman/electric field and the skyrmion charge, respectively.}
\vspace{0.1in}
\begin{tabular}{|c|c|c|c|c|c|}\hline
&\bt{Magnetic Phase}      &\bt{Rot. Symmetry}   &\bt{Induced charge order}     &\bt{Magnetoelectricity}             &$\bm{{\cal B}}$\bt{-induced} $\bm{{\cal C}}$\\\hline
$\bm{1.}$&\bt{IC-MS}($\bm{Q}_1$)           &\bt{C}$_{\bm{2}}$  & \{$\pm2\bm{Q}_1$\}                                          &\oust                           &\oust\\\hline
$\bm{2.}$&\bt{IC-CSDW}                     &\bt{C}$_{\bm{4}}$  & \{$\pm2\bm{Q}_1$, $\pm2\bm{Q}_2$, $\pm\bm{Q}_1\pm\bm{Q}_2$\}&\oust                           &\oust\\\hline
$\bm{3.}$&\bt{IC-SVC}                      &\bt{C}$_{\bm{4}}$  & \{$\pm2\bm{Q}_1$, $\pm2\bm{Q}_2$\}                          &\oust                           &\oust\\\hline
$\bm{4.}$&\bt{MH}($\bm{Q}_2$)              &\bt{C}$_{\bm{2}}$  & \oust                                    &\yes ($\bm{M}_{\bm{0}}\propto\bm{Q}_2\cdot\bm{{\cal E}}$)&\oust\\\hline
$\bm{5.}$&\bt{MS}($\bm{Q}_1$)$||$\bt{MH}($\bm{Q}_2$)&\bt{C}$_{\bm{2}}$& $\{\pm2\bm{Q}_1\}$, $\{{\color{gray}\pm2\bm{Q}_2}\}$, $\{\pm\bm{Q}_1\pm\bm{Q}_2\}$&\yes ($\bm{M}_{\bm{0}}\propto\bm{Q}_2\cdot\bm{{\cal E}}$)&\oust\\\hline
$\bm{6.}$&\bt{MS}($\bm{Q}_1$)$\perp$\bt{MH}($\bm{Q}_2$)&\bt{C}$_{\bm{2}}$& \{$\pm2\bm{Q}_1$\}                    &\yes ($\bm{M}_{\bm{0}}\propto\bm{Q}_2\cdot\bm{{\cal E}}$)&\oust\\\hline
$\bm{7.}$&\bt{DPMH}($\hat{\bm{n}}_1=\hat{\bm{n}}_2$)&\bt{C}$_{\bm{2}}$ & \{$\pm(\bm{Q}_1-\bm{Q}_2)$\} &\yes $(\bm{M}_{\bm{0}}\propto(\bm{Q}_1+\bm{Q}_2)\cdot\bm{{\cal E}})$ &\oust\\\hline
$\bm{8.}$&\bt{SWC}$_{\bm{4}}$        &\bt{C}$_{\bm{4}}$ & \{$\pm2\bm{Q}_1$, $\pm2\bm{Q}_2$\}, \{$\pm\bm{Q}_1\pm\bm{Q}_2$\}     &\yes                      &\yes(${\cal C}=\pm1$)\\\hline
$\bm{9.}$&\bt{SWC}$_{\bm{2}}$($\hat{\bm{n}}_{1,2}$ of Eq.\eqref{eq:MW2}) &\bt{C}$_{\bm{2}}$ & \{$\pm2\bm{Q}_2$\}, \{$\pm\bm{Q}_1\pm\bm{Q}_2$\}&\yes       &\yes(${\cal C}=\pm1$)\\\hline
\end{tabular}
\label{table:ICMagneticSignatures}
\end{table*}

While the IC scenario provides a number of candidates for the unidentified magnetic phase appearing in Na-doped BaFe$_2$As$_2$, an unambiguous identification requires additional experimental measurements. In order to identify and distinguish these nine magnetic ground states and their possible presence in FeSCs, one can employ a combination of experimental techniques capable of resolving the magnetically induced charge order in the system and the emergence of a non-zero magnetoelectric coupling. The former can be detected in X-ray scatte\-ring and the emerging Bragg peaks. For the latter, one can measure the magnetization ge\-ne\-ra\-ted by applying an external electric field ($\bm{{\cal E}}$) or via indu\-cing a current ($\bm{I}$) flow. We predict that a homogeneous magnetization can appear in the presence of helices or whirls in the magnetic ground state, depen\-ding on the orientation of $\bm{{\cal E}}$ or $\bm{I}$. Our results regarding the experimental signatures of the nine magnetic phases are summarized in Table~\ref{table:ICMagneticSignatures}. 

By adopting two five-orbital models~\cite{ikeda10,graser10} based on density functional theory (DFT) calculations for BaFe$_2$As$_2$ and LaFeAsO, respectively, we demonstrate that IC magnetism is a rea\-listic scenario for the fa\-mi\-ly of FeSCs. Remarkably, textured magnetic phases are accessible for both bandstructure types considered here. In fact, BaFe$_2$As$_2$ supports a variety of textured phases. On the other hand, textured phases can be stabilized in LaFeAsO by taking into account orbital selectivity~\cite{OrbitalSelectivityLuca,OrbitalSelectivityYi,OrbitalSelectivityScience,OrbitalSelectivityPRB}, mo\-del\-led in the present case with orbital-dependent interactions. For the explored parameter regime we find that a C$_4$-symmetric spin-whirl crystal phase becomes favored in both cases. Remarkably, this can acquire a topologically non-trivial skyrmion charge by the application of a weak Zeeman field ($\bm{{\cal B}}$).

The possible emergence of new magnetic phases in FeSCs does not only add to the established commensurate picture regarding the magnetic phase diagram, it also opens up perspectives for crafting novel to\-po\-lo\-gi\-cal phases. In fact, the experimentally confirmed microscopic coexistence of magnetism and supercon\-duc\-ti\-vi\-ty in FeSCs~\cite{avci14a,wang16a,ni08a,nandi10a} appears as a promising platform for rea\-li\-zing intrinsic TSCs. While symmetry constraints~\cite{KotetesClassi} prohibit crafting \textit{strong} TSCs~\cite{QiZhang} by solely com\-bi\-ning the standard commensurate magnetic phases with spin-singlet superconductivity, the possible emergence of non-coplanar magnetic textures could provide a way around this obstacle, as previous pro\-po\-sals on artificial chiral topological SCs~\cite{KotetesClassi,Nakosai2013,Ojanen,Sedlmayr,Mendler,Chen,Loss} suggest. We note that the magnetically-engineered TSCs relevant here, are distinct from standard SOC-generated TSCs which may also be relevant in some of the FeSCs, as revealed by recent angle resolved photoemission spectroscopy (ARPES) and scanning tunneling microscopy (STM) experiments~\cite{hongding1,hongding2}.

The content of this paper is organized as follows: In Sec.~\ref{Sec:IILandau} we present the general Landau formalism and discuss the emergence of nine IC magnetic phases, among which one finds a number of potential candidates for the recently discovered phase in Ba$_{1-x}$Na$_x$Fe$_{2}$As$_{2}$~\cite{wang16a}. The detailed magnetic structure of these nine phases are presented in Sec.~\ref{Sec:ICMagneticPhases}. In Sec.~\ref{Sec:GenPhaseDiagram} we show the generic magnetic phase diagram, illustrating the leading magnetic instability as a function of the Landau parameters. Sec.~\ref{Sec:expsignatures} discusses experimental signatures of the IC phases, focussing on how to distinguish them from one another without relying on e.g. detailed polarized neutron scattering measurements. In Sec.~\ref{sec:pnictide_section} we turn to FeSCs and derive the Landau functional for the two microscopic five-band models, and include interactions in a mean-field manner using a standard multi-orbital Hubbard-Hund Hamiltonian. This allows for a microscopic determination of the Landau coefficients, the thermodynamic magnetic phase diagram, and subsequently, an analysis of the energetics of the IC magnetic phases. Moreover, we discuss the effects of the inversion-symmetric atomic SOC on the magnetic phase diagram and the appea\-ran\-ce of non-coplanar magnetic textures. Finally, Sec.~\ref{Sec:Conclusions} summarizes our results and discusses further directions.

\section{Landau formalism for incommensurate magnetism}\label{Sec:IILandau}

To identify the accessible IC magnetic phases for a generic itinerant system with tetragonal symmetry and ordering wavevectors $\bm{Q}_{1,2}=(\pi-\delta,0)/(0,\pi-\delta)$ or $\bm{Q}_{1,2}=(\pi,\delta)/(-\delta,\pi)$, we write down the Landau fun\-ctional up to quartic order with respect to the magnetic order parameters $\bm{M}(\bm{Q}_{1,2}) \equiv \bm{M}_{1,2}$. Contrary to the commensurate case ($\delta=0$) where $\bm{M}_{1,2}=\bm{M}_{1,2}^*$, in the present IC situation the magnetic order parameters are complex since $\bm{M}(\bm{Q}_{1,2})\neq\bm{M}(\bm{Q}_{1,2})^*\equiv\bm{M}(-\bm{Q}_{1,2})$. This results in the following expression for the Landau free ener\-gy func\-tional:
\bea
F&=&\alpha(|\bm{M}_1|^2+|\bm{M}_2|^2)+\frac{\tilde{\beta}}{2}(|\bm{M}_1|^2+|\bm{M}_2|^2)^2\nonumber\\ 
&+&\frac{\beta-\tilde{\beta}}{2}(|\bm{M}_1^2|^2+|\bm{M}_2^2|^2)+(g-\tilde{\beta})|\bm{M}_1|^2|\bm{M}_2|^2\nonumber\\ 
&+&\frac{\tilde{g}}{2}(|\bm{M}_1\cdot\bm{M}_2|^2+|\bm{M}_1\cdot\bm{M}_2^{\ast}|^2)\label{eq:free_energy}\,.
\eea
The above was previously studied by Schulz~\cite{schulz90}, restricted however, to the possible occurrence of IC magnetism in high-$T_{\rm c}$ cuprates. We note that the free energy given in Eq.~(\ref{eq:free_energy}) is obtainable from a microscopic electronic model, as will be demonstrated in Sec.~\ref{sec:pnictide_section} below.

The Landau functional in Eq.~\eqref{eq:free_energy} is invariant under complex-conjugation (${\cal K}$), time-reversal (${\cal T}$), D$_{\rm 4h}$ point group ope\-ra\-tions, SO(3) spin rotations and translations ($t_{\bm{a}}$, with $\bm{a}$ the direct lattice shift vector). At quadratic level the above symmetry becomes artificially enhanced yielding a de\-ge\-ne\-ra\-cy among the possible candidate phases for the leading magnetic instability that sets in when $\alpha<0$. The latter guides us to pa\-ra\-me\-trize the order parameters by
\bea 
\bm{M}_{1}=M\cos\eta\ph\hat{\bm{n}}_1\quad{\rm and}\quad\bm{M}_{2}=M\sin\eta\ph\hat{\bm{n}}_2,
\eea 

\noi with $|\hat{\bm{n}}_{1,2}|^2=1$ and $\eta\in[0,\pi/2]$. Note that the complex spin-vectors generally satisfy $|\hat{\bm{n}}_{1,2}^2|\leq1$. Translational invariance allows us to arbitrarily and independently choose the overall phase of the vectors $\hat{\bm{n}}_{1,2}$. On the other hand, spin-rotational invariance allows for further simplifications, for instance setting ${\rm Re}[\hat{\bm{n}}_1]$ parallel to the $z$ spin axis. Note that this is not possible if SOC is present since this introduces anisotropy in spin-space. The effects of a weak SOC will be discussed in Sec.~\ref{Sec:SOC}.

Under these conditions, extremizing the Landau func\-tional with respect to $\eta$ yields
\bea
\sin(2\eta)=0,\,\,
\cos(2\eta)=\frac{|\hat{\bm{n}}_1^2|^2-|\hat{\bm{n}}_2^2|^2}{2G+2\tilde{G}P-(|\hat{\bm{n}}_1^2|^2+|\hat{\bm{n}}_2^2|^2)}\,,\quad\label{eq:Eta}
\eea

\noi where we have introduced
\bea
G\equiv\frac{g-\tilde{\beta}}{\beta-\tilde{\beta}}\,,\tilde{G}\equiv\frac{\tilde{g}}{\beta-\tilde{\beta}}\,, 
P\equiv\frac{|\hat{\bm{n}}_1\cdot\hat{\bm{n}}_2|^2+|\hat{\bm{n}}_1\cdot\hat{\bm{n}}_2^*|^2}{2}\,.\quad
\eea

\noi For $\sin(2\eta)=0$ we retrieve single-$\bm{Q}$ phases since $\eta=0$ ($\eta=\pi/2$) implies that only the order parameter with wavevector $\bm{Q}_1$ ($\bm{Q}_2$) appears. 

The remaining extrema arise for values of $\eta$ determined by the $\cos(2\eta)$, leading to double-$\bm{Q}$ phases. For $|\hat{\bm{n}}_1^2|\neq|\hat{\bm{n}}_2^2|$ we have $\eta\neq\pi/4$. By observing that $\tan\eta=|\bm{M}_2|/|\bm{M}_1|$ we obtain that in this case $|\bm{M}_1|\neq|\bm{M}_2|$ and, thus, all the arising double-$\bm{Q}$ phases violate C$_4$-symmetry leaving only a C$_2$ subgroup intact. If instead $|\hat{\bm{n}}_1^2|=|\hat{\bm{n}}_2^2|$, not ne\-ces\-sa\-ri\-ly all double-$\bm{Q}$ phases are C$_4$-symmetry violating. After excluding special or singular values of the Landau coefficients we find nine extrema of the free energy. These are presented in Table~\ref{table:ICMagneticSignatures} and in Sec.~\ref{Sec:ICMagneticPhases}. Further details regarding the corresponding free energy are presented in Appendix~\ref{App:MagPhases}. Finally, note that the two possible values for the sign of $\beta-\tilde{\beta}$ give rise to two generic phase diagrams that will be presented in detail in Sec.~\ref{Sec:GenPhaseDiagram}.

\section{Incommensurate magnetic phases}\label{Sec:ICMagneticPhases}

The order pa\-ra\-me\-ters of the accessible magnetic phases are identified by the values of $\eta$ and the spin-vectors $\hat{\bm{n}}_{1,2}$, which are determined by mi\-ni\-mi\-zing the free ener\-gy of Eq.~\eqref{eq:free_energy}. In addition to the IC extensions of the three well-known commensurate single-$\bm{Q}$ stripe (MS) and double-$\bm{Q}$, collinear (CSDW) and non-collinear (SVC) phases, we uncover a single-$\bm{Q}$ helix phase along with five further double-$\bm{Q}$ phases. These double-$\bm{Q}$ phases include a number of C$_4$-symmetry breaking phases.

\subsection{IC extensions of the known commensurate magnetic phases}\label{Sec:ICExte}

We commence by considering the IC generalizations of the three well-studied commensurate phases, the C$_2$-symmetric MS phase along with the C$_4$-symmetric CSDW and SVC phases.\newline

{\bf IC Magnetic Stripe (IC-MS):} This constitutes the generalization of the typical magnetic $\rm{C}_2$-symmetric phase, prevalent in FeSCs. As a single-$\bm{Q}$ phase, it can appear for either $\bm{Q}_1$ or $\bm{Q}_2$. Since the latter are equi\-va\-lent, they can lead to domains in actual materials. Throughout this work we focus on single-domain realizations and choose the magnetic ordering vector of the IC-MS to be $\bm{Q}_1=(\pi-\delta,0)$ throughout. Thus we obtain
\bea
\hat{\bm{n}}_1=(0,0,1)\quad{\rm and}\quad\hat{\bm{n}}_2=(0,0,0)\,,
\eea

\noi that yields the magnetic order depicted in Fig.~\ref{fig:IC_generalizations}(a).\newline

{\bf IC Charge-Spin Density Wave (IC-CSDW):} This is a double-$\bm{Q}$ C$_4$-symmetric phase which constitutes the IC extension of the CSDW phase and can be seen as two superimposed IC-MS phases with parallel magnetic order parameter spin-vectors. An example of this magnetic order with $\bm{Q}_{1,2}=(\pi-\delta,0)/(0,\pi-\delta)$ and
\bea
\hat{\bm{n}}_{1}=\left(0\,,0\,,1\right)\quad{\rm and}\quad\hat{\bm{n}}_{2}=\left(0\,,0\,,1\right)\,,
\eea
is shown in Fig.~\ref{fig:IC_generalizations}(b).\newline

{\bf IC Spin-Vortex Crystal (IC-SVC):} This ge\-ne\-ra\-lizes the commensurate SVC phase. It is a double-$\bm{Q}$ C$_4$-symmetric phase and similarly to the IC-CSDW above, it can be viewed as two MS phases superimposed. The difference in this case is that the spin-vectors of the magnetic order parameter are orien\-ted in a perpen\-di\-cu\-lar fashion. An example of such a magnetic order with $\bm{Q}_{1,2}=(\pi-\delta,0)/(0,\pi-\delta)$ is
\bea
\hat{\bm{n}}_{1}=\left(0\,,1\,,0\right)\quad{\rm and}\quad\hat{\bm{n}}_{2}=\left(1\,,0\,,0\right)\,.\label{eq:ICSVCReal}
\eea
This phase is depicted in Fig.~\ref{fig:IC_generalizations}(c). It is worth commen\-ting on the nomenclature of the particular magnetic phase, which reflects the presence of singular points in the spatial profile of the magnetization. The singular points act as sources of vorticity, cf. Refs.~\onlinecite{Volovik,Graphene}.
This becomes apparent by rewriting the here two-dimensional magnetization vector as follows $\bm{M}(\bm{r})=(M_x(\bm{r}),M_y(\bm{r}),0)\equiv|\bm{M}(\bm{r})|(\cos\varphi(\bm{r}),\sin\varphi(\bm{r}),0)$. The vorticity sources are located at points in space for which $|\bm{M}(\bm{r})|=0$ and for which at the same time the phase field, $\varphi(\bm{r})$, jumps by $n$ multiples of $2\pi$ as we move around a loop encircling the singularity. Here $n$ corresponds to the number of vorticity units carried by the defect.
For the choice of spin-vectors in Eq.~\eqref{eq:ICSVCReal}, the coordinates for the vorti\-ci\-ty sources are given by $\bm{r}_s=\frac{\pi}{2Q}(\pm1,\pm1)$. The vorticity points are also depicted in Fig.~\ref{fig:IC_generalizations}(c), where we have chosen the unit cell such that it contains one of the vortices located at the center. One finds that each singular point contributes with a single unit of vorticity, while within a single magnetic unit cell the total vorticity is zero.

\subsection{New magnetic phases arising by virtue of incommensurability}

Apart from the anticipated IC extensions of the three well known commensurate magnetic phases encountered in the FeSCs, we find additional phases unique to the IC case. These are portrayed in Fig.~\ref{fig:IC_magnetic_phases}(a)-(f). In (a) we present a single-$\bm{Q}$ magnetic helix (MH) phase. In (b)/(c) the MH phase coexists with an in-plane/out-of-plane magnetic spiral. In (d) a C$_2$-symmetric double parallel MH (DPMH) phase is shown, while in (e) a non-coplanar C$_4$-symmetric spin-whirl crystal (SWC$_{4}$) phase is depicted. A C$_2$ version (SWC$_2$) of this phase is shown in (f).\newline

{\bf Magnetic Helix (MH):} This is a single-$\bm{Q}$ C$_2$-symmetric phase. As we are in a position to arbi\-tra\-ri\-ly choose the ordering wavevector, below we present the spin-vector structure for the $\bm{Q}_1$ ordering wavevector  
\bea
\hat{\bm{n}}_1 = \frac{1}{\sqrt{2}}(i,0,1) \quad {\rm and} \quad \hat{\bm{n}}_2=(0,0,0)\,.\label{eq:mh_config}
\eea

\noi For $\bm{Q}_1 = (\pi-\delta,0)$ the magnetic texture is depicted in Fig.~\ref{fig:IC_magnetic_phases}(a). Note that this is the only IC magnetic phase for which the modulus of the magnetization in coordinate space, $|\bm{M}(\bm{r})|$, is spatially constant.\newline 

{\bf Magnetic Stripe with parallel Magnetic Helix (MS$||$MH):} This double-$\bm{Q}$ C$_2$-symmetric phase consists of a MH with ordering wavevector $\bm{Q}_{1(2)}$  and an IC-MS with wavevector $\bm{Q}_{2(1)}$. The magnetic moment ari\-sing from the MS is oriented inside the winding plane of the MH. By choosing one of the above allowed and equivalent wavevector configurations, the spin-vectors are given by
\bea
\hat{\bm{n}}_{1}=\left(0\,,0\,,1\right)\phd{\rm and}\phd\hat{\bm{n}}_{2}=\left(i\sin\lambda\,,0\,,\cos\lambda\right)\,,
\eea

\noi and the magnetic order is shown in Fig.~\ref{fig:IC_magnetic_phases}(b) for $\bm{Q}_{1,2}=(\pi-\delta,0)/(0,\pi-\delta)$. Note that the parameter $\lambda$ is provided by minimizing the free energy and its expression is presented in Appendix~\ref{App:MagPhases}.\newline

{\bf Magnetic Stripe with perpendicular Magnetic Helix (MS$\perp$MH):} This double-$\bm{Q}$ C$_2$-symmetric phase is composed by a MH and an IC-MS with magnetic moment of the stripe oriented out-of-the winding plane defined by the MH. The respective magnetic vectors read
\bea
\hat{\bm{n}}_{1}=(0,0,1)\phd{\rm and}\phd\hat{\bm{n}}_{2}=\frac{1}{\sqrt{2}}\left(i\,,1\,,0\right)\,,
\eea
and for $\bm{Q}_{1,2}=(\pi-\delta,0)/(0,\pi-\delta)$ the order is depicted in Fig.~\ref{fig:IC_magnetic_phases}(c). Note that this is an example of a non-coplanar magnetic order.\newline

{\bf Double Parallel Magnetic Helix (DPMH):} This is a double-$\bm{Q}$ coplanar C$_2$-symmetric phase (in spite of $\eta=\pi/4$) consisting of two helices ordered at $\bm{Q}_{1,2}$, with spin winding within the same spin plane. This phase is described by identical spin-vectors
\bea
\hat{\bm{n}}_{1,2}=\frac{1}{\sqrt{2}}\left(i\,,0\,,1\right)\,.
\eea

\noi For $\bm{Q}_{1,2}=(\pi-\delta,0)/(0,\pi-\delta)$ the order is shown in Fig.~\ref{fig:IC_magnetic_phases}(d). \newline

{\bf C$_4$-symmetric Spin-Whirl Crystal (SWC$_4$):} This is a double-$\bm{Q}$ C$_4$-symmetric textured phase with
\bea
&&\hat{\bm{n}}_{1}=\left(i\cos\lambda\,,0\,,\sin\lambda\right)\,,\phd\phd\no\\
&&\hat{\bm{n}}_{2}=\left(0\,,i\cos\lambda\,,\sin\lambda\right)\,.\label{eq:MW4}
\eea
For $\bm{Q}_{1,2}=(\pi-\delta,0)/(0,\pi-\delta)$ we obtain the coordinate space magnetization profile:
\bea
\bm{M}(\bm{r}) &=&\sqrt{2}M\begin{pmatrix}
\cos \lambda \sin \left(\bm{Q}_1 \cdot \bm{r}\right) \\
\cos \lambda \sin \left(\bm{Q}_2 \cdot \bm{r}\right) \\
\sin\lambda \left[\cos \left(\bm{Q}_1 \cdot \bm{r} \right)  + \cos \left(\bm{Q}_2 \cdot \bm{r}\right)\right]  
	\end{pmatrix}\,,\qquad\label{eq:SWCprofile}
\eea

\noi plotted in Fig.~\ref{fig:IC_magnetic_phases}(e) for a specific choice of $\lambda$, whose expression is given in Appendix~\ref{App:MagPhases}.

Similar to the IC-SVC phase, also the present spatial profile of the magnetization contains singular points ac\-ting as sources of topological charge located at $\bm{r}=\bm{r}_s$. The difference is that, in contrast to the IC-SVC case, here the magnetization is a three-dimensional vector in spin space. Thus, instead of vorticity which characterizes singularities of a coplanar magnetic texture, we can employ the singular Berry curvature~\cite{Murakami} ($\Omega_{xy}^s(\bm{r})=\pi\delta(\bm{r}-\bm{r}_s)$) or $\pi$-Berry flux~\cite{Volovik,Graphene,TIexp,Fu} for assigning topological charge to the point singularities for a non-coplanar magnetix texture~\cite{Fu}. This explains the reasoning behind naming the particular magnetic textures as spin-whirls. Note that if we focus on the vicinity of the singular points, the vorticity and the $\pi$-Berry flux coin\-cide since the texture becomes effectively coplanar~\cite{Warped}. Importantly, one should not confuse the spin-whirls discussed here with spin-skyrmions~\cite{Volovik}. In the latter case the magnetic profile contains no point singularities, the Berry curvature~\cite{Niu} $\Omega_{xy}(\bm{r})$ can be defined for all $\bm{r}$ and the skyrmion charge is non-zero. For further details see Sec.~\ref{Sec:Skyrmions}.

For the profile of Eq.~\eqref{eq:SWCprofile}, the singular points are located at $(\pi/Q,0)$ and $(0,\pi/Q)$, as given by the zeros of the modulus $|\bm{M}(\bm{r})|$. Each one of the two non-equivalent so-called nodal points, contributes with a unit of topological charge. These nodes can be gapped out by either violating the C$_4$ symmetry or the antiunitary symmetry $\Theta={\cal T}t_{(\pi/Q,\pi/Q)}$, consisting of the successive ope\-ra\-tion of time reversal ${\cal T}$ and a direct-lattice translation by $(\pi/Q,\pi/Q)$ (see also Ref.~\onlinecite{SkyrmionComment}). We find $\Theta^2=-{\rm 1}$ which implies that nodal points of unit topological charge come in pairs, or, that the skyrmion charge is zero if the nodes are gapped out by breaking C$_4$ symmetry, cf. the next paragraph. Any type of a bulk perturbation that does not respect the above $\Theta$ symmetry lifts these nodes, thus, resulting in a non-zero Berry curvature in the entire magnetic unit cell. Depending on the perturbation, the latter symmetry breaking can endow the SWC with non-zero skyrmion charge as discussed in Sec.~\ref{Sec:Skyrmions}.\newline

{\bf C$_2$-symmetric Spin-Whirl Crystal (SWC$_2$):} This phase constitutes a two-fold symmetric version of the one above. It consists of an isotropic helix for $\bm{Q}_{1,(2)}$ coe\-xi\-sting with an anisotropic helix for $\bm{Q}_{2,(1)}$, i.e.
\bea
\hat{\bm{n}}_{1}=\frac{1}{\sqrt{2}}\left(i\,,0\,,1\right)\ph{\rm and}\ph\hat{\bm{n}}_{2}=\left(0\,,i\sin\lambda\,,\cos\lambda\right)\,,\quad\label{eq:MW2}
\eea

\noi which for a choice of $\lambda$ is depicted in Fig.~\ref{fig:IC_magnetic_phases}(f). For the same type of IC magnetic ordering, as discussed for SWC$_4$, we find that the nodal points are gapped out due to the broken C$_4$ symmetry. Nevertheless, the $\Theta$ symmetry is still present and leads to a zero skyrmion charge.

\section{Generic Magnetic phase diagrams}\label{Sec:GenPhaseDiagram} 

\begin{figure}[b!]
\centering
\includegraphics[width=\columnwidth]{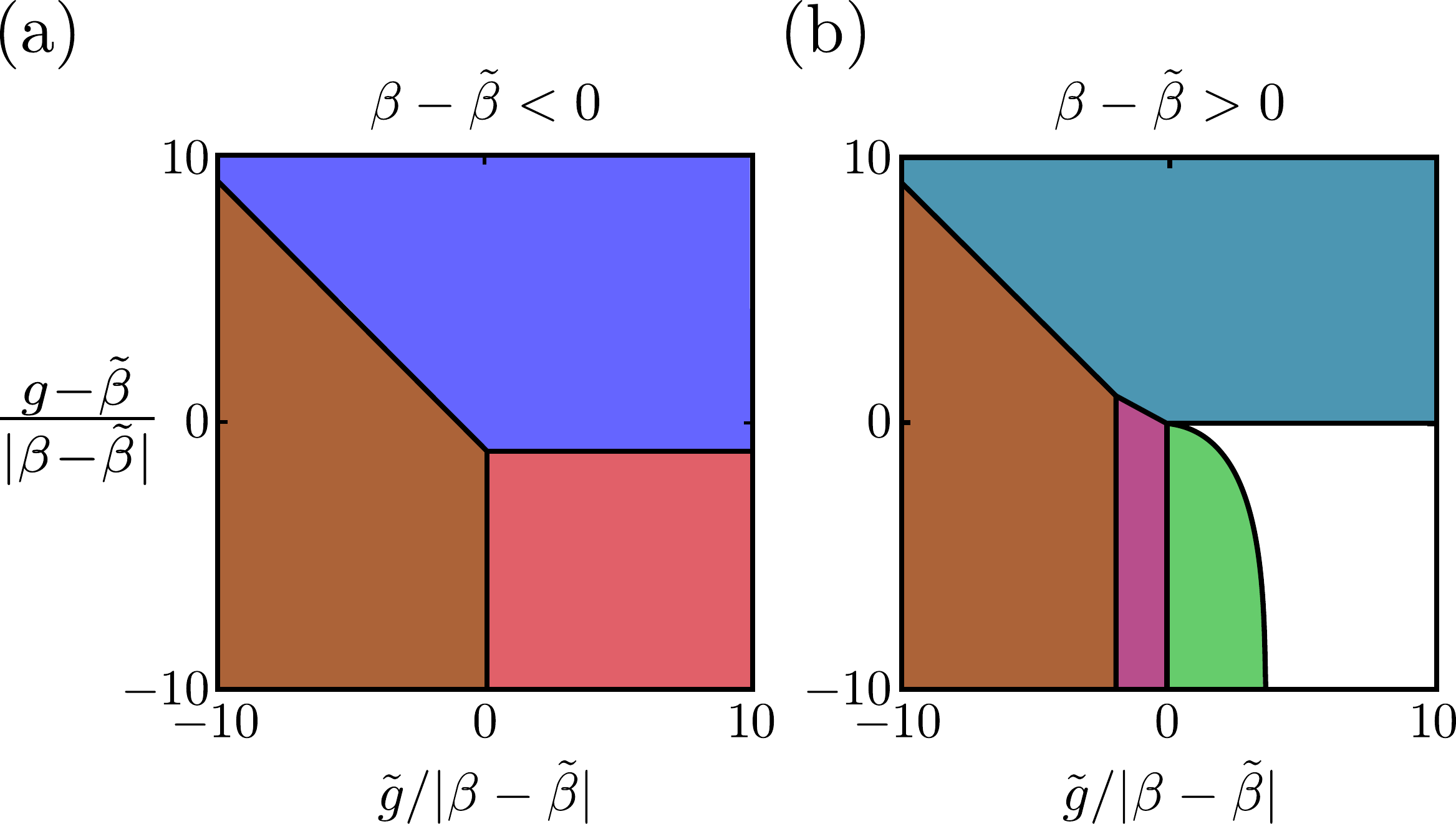}
\caption[]{\label{fig:phase_diagram_G_Gtilde} Phase diagrams for the generic Landau func\-tional in Eq.~(\ref{eq:free_energy}) with: C$_2$-symmetric IC magnetic stripe (IC-MS) (\tikz\draw[inc_stripe,fill=inc_stripe] (0,0) circle (.5ex);), C$_4$-symmetric IC charge-spin density wave (IC-CSDW) phase (\tikz\draw[collinear,fill=collinear] (0,0) circle (.5ex);), C$_4$-symmetric IC spin-vortex crystal (IC-SVC) phase (\tikz\draw[non_collinear,fill=non_collinear] (0,0) circle (.5ex);), C$_2$-symmetric magnetic helix (MH) (\tikz\draw[mag_spiral,fill=mag_spiral] (0,0) circle (.5ex);), C$_2$-symmetric IC magnetic stripe with a perpendicular ($\perp$) magnetic helix (MS$\perp$MH) phase (\tikz\draw[black,fill=spiral_w_orth_stripe] (0,0) circle (.5ex);), C$_2$-symmetric double parallel magnetic helix (DPMH) (\tikz\draw[coplanar,fill=coplanar] (0,0) circle (.5ex);), C$_4$-symmetric spin-whirl crystal (SWC$_4$) phase (\tikz\draw[c4_noncoplanar,fill=c4_noncoplanar] (0,0) circle (.5ex);). The IC extensions of the original three phases (\tikz\draw[inc_stripe,fill=inc_stripe] (0,0) circle (.5ex);,\tikz\draw[collinear,fill=collinear] (0,0) circle (.5ex);,\tikz\draw[non_collinear,fill=non_collinear] (0,0) circle (.5ex);) are the only ones present when $\beta-\tilde{\beta}<0$, however, for $\beta-\tilde{\beta}>0$ four of the new phases occupy a substantial region of the phase diagram.}
\label{fig:GenPhaseDiagram}
\end{figure}

Having presented the allowed leading magnetic instabilities and related order parameters of the generic Landau theo\-ry of Eq.~\eqref{eq:free_energy}, we now show the generic phase dia\-gram with respect to the rescaled parameters $g-\tilde{\beta}$ and $\tilde{g}$. We consider two cases, depending on the sign of $\beta-\tilde{\beta}$. The resulting phase diagrams are presented in Fig.~\ref{fig:phase_diagram_G_Gtilde}. The $\beta - \tilde{\beta}<0$ case is very reminiscent of the commensurate phase diagram, while the case with $\beta - \tilde{\beta}>0$ exhibits four out of the six new phases alongside the IC-CSDW. In general, we find that the MH, IC-MS and IC-CSDW phases dominate in a large part of the phase diagrams.

For $\beta-\tilde{\beta}<0$ only the three IC extensions of the commensurate phases appear. This resembles the well-known commensurate phase diagram of Refs.~\onlinecite{lorenzana08,wang15,christensen17}. For $\beta-\tilde{\beta}>0$ the MH appears for a wide range of values of the Landau parameters. In the lower right quadrant we find both the MS$\perp$MH phase and the SWC$_4$ phase, while the DPMH phase is confined to a thin sliver for negative $g-\tilde{\beta}$ and $\tilde{g}$. The remaining part of the phase diagram is occupied by the IC-CSDW phase. 

Comparing the two phase diagrams reveals that the IC-CSDW occupies almost the same region in both, while the IC-MS is replaced by the MH as $\beta-\tilde{\beta}$ changes sign. For $\beta-\tilde{\beta}>0$, the DPMH and SWC$_4$ phases share a common boundary defined by the critical line $\tilde{g}/|\beta-\tilde{\beta}|=0$, where a degeneracy among phases with an arbitrary re\-la\-ti\-ve orientation of the spin-vectors $\bm{n}_{1,2}$ occurs. In the lower right quadrant, the MS$\perp$MH phase replaces the IC-SVC as $\beta-\tilde{\beta}$ becomes positive. One further observes a certain degree of similarity between the two phase dia\-grams. Since the phase diagram for $\beta-\tilde{\beta}<0$ resembles the commensurate one, it is useful to consi\-de\-r the commensurate limit $(\pi,0)$ and $(0,\pi)$ of the IC magnetic wavevectors $\bm{Q}_{1,2}$, for the textured phases. One obtains MH$\rightarrow$MS, MS$\perp$MH$\rightarrow$SVC, DPMH$\rightarrow$CSDW and SWC$_{4}\rightarrow$SVC~\cite{CommensurateLimit}. The latter limits provide a clear hint regarding the expected position of each textured phase in Fig.~\ref{fig:phase_diagram_G_Gtilde}(b), by virtue of its continuity to the respective commensurate order.

The SWC$_2$ and MS$\parallel$MH do not appear as free energy minima near the paramagnetic-magnetic transition in the parameter regime studied here. Based on the structure of the generic phase diagrams, we can conjecture that this holds in the entire parameter space. However, this statement is challenging to prove using the analytical expressions discussed in Appendix~\ref{App:MagPhases}. Therefore, we will still consider the appearance of these phases as a possible scenario. A reason to leave such a possibility open is that, since they constitute extrema of the free energy, they may still appear as metastable phases. Even more, these two phases may become thermodynamically stable under the additional presence of external fields or other more general conditions not taken into account in this study.

Finally, we remark that for the values of $g$, $\tilde{g}$, $\beta$ and $\tilde{\beta}$ considered here, the free energy is bounded from below and we can truncate the free energy expansion at quartic order. Otherwise, one should include appropriate contributions from higher order terms.

\section{Signatures and experimental detection}\label{Sec:expsignatures}

From the previous section it is clear that several IC magnetic phases may become accessible, and therefore it is desirable to identify ways to differentiate them experimentally. While directly probing the spatial profile of the magnetization using spin-resolved means can certainly uniquely identify them, it can be a formidable experimental task. Thus, it is important to elaborate on alternative detection methods that do not rely on the texture's profile alone, but can still provide an unambiguous identification of the stabilized magnetic order. For this purpose we consider the combination of charge and magnetoelectric measurements. Below we provide more details regarding these two, while we additionally discuss further interesting features of the SWC$_{4,2}$ phases, i.e. that they can be converted into spin-skyrmion crystal phases via the application of an external magnetic field.  

\subsection{Induced charge order}

A common characteristic of almost all the IC magnetic phases is that the respective modulus of the magnetization profile $|\bm{M}(\bm{r})|$ is not constant, but rather spatially varying. The only exception is the MH phase. As a consequence of this spatial inhomogeneity, charge order is induced when IC magnetism occurs. For a ge\-ne\-ral magnetic profile with Fourier decomposition $\bm{M}(\bm{r})=\sum_{\bm{q}}\bm{M}_{\bm{q}}e^{i\bm{q}\cdot\bm{r}}$, the induced charge order $\rho(\bm{r})=\sum_{\bm{q}}\rho_{\bm{q}}e^{i\bm{q}\cdot\bm{r}}$ has Fourier components proportional to the scalar product of the magnetic order parameters' spin-vectors ($\bm{M}_{\bm{q}}=|\bm{M}_{\bm{q}}|\hat{\bm{n}}_{\bm{q}}$):
\bea
\rho_{\bm{q}+\bm{p}}\propto|\bm{M}_{\bm{q}}||\bm{M}_{\bm{p}}|\hat{\bm{n}}_{\bm{q}}\cdot\hat{\bm{n}}_{\bm{p}}\,.\label{eq:InducedCharge}
\eea

\noi Note that the above are the components generated at lowest order with respect to the magnetic order para\-me\-ters, while contributions of higher order will lead to weaker effects (see also Ref.~\onlinecite{Tsunetsugu}). Given that here the IC magnetic ordering is considered to occur at $\bm{Q}_{1,2}$, Eq.~\eqref{eq:InducedCharge} implies that there exists a set of eight possible induced components at lowest order, $\rho_{\bm{q}}$, with $\bm{q}=\{\pm2\bm{Q}_1,\pm2\bm{Q}_2,\pm\bm{Q}_1\pm\bm{Q}_2\}$. The magnetic-charge order coupling, leading to Eq.~\eqref{eq:InducedCharge}, is derived in Appendix~\ref{App:AppendixInducedCharge}.

Based on Sec.~\ref{Sec:ICMagneticPhases} we can directly infer the Fourier components of the induced charge order for the dif\-fe\-rent cases. In Table~\ref{table:ICMagneticSignatures} we display the resulting induced charge density components depending on the magnetic order. First of all, one observes that the MH, IC-SVC and DPMH phases can be unambiguously identified via solely using the electronic charge density's Bragg peaks. The remaining six phases are split into three subsets: \{IC-MS,\,MS$\perp$MH\}, \{IC-CSDW,\,SWC$_4$\}, and \{MS$||$MH,\,SWC$_2$\}. The magnetic phases belonging to the same subset are precisely or practically indistingui\-sha\-ble via this particular experimental approach. However, the magnetic phases of each subset can be discerned based on their different magnetoelectric properties, which are discussed in the next subsection. 

Before concluding this subsection, we point out that detecting charge order is more suitable for identifying IC, rather than commensurate, magnetic orders in FeSCs. This is because both commensurate magnetic ordering wavevectors $(\pi,0)$ and $(0,\pi)$, defined in the 1Fe/unit cell, translate to $(\pi,\pi)$ in the 2Fe/unit cell. In the latter, physical, unit cell, commensurate magnetism induces a charge order Bragg peak at $(0,0)$, which coincides with a lattice Bragg peak. However, for IC magnetism the charge order is also IC and the peaks are shifted away from $(0,0)$ and should, in principle, be detectable.

\subsection{Magnetoelectric effects}

The textured nature of the magnetic phases accessible through incommensurability opens up the pos\-si\-bi\-li\-ty of a finite magnetoelectric coupling.
This is linked to magnetization profiles in which the spin moment smoothly winds while sweeping particular paths in coor\-dinate space. Notably, such exotic magnetoelectric manifestations are incompatible with commensurate magnetic phases and their related IC extensions. 

To be more specific regarding the type of magnetoelectric phenomena to be explored in the present work, we remind the reader that all the six new IC phases di\-sco\-ve\-red here consist of a single- or double-$\bm{Q}$ MH confi\-gu\-ra\-tions. However, it is well known from other studies that a MH is equivalent to a SOC term in the presence of an orthogonally oriented ferromagnetic field~\cite{Braunecker,Ivar,Flensberg}. To make the latter statement transparent, consider the MH Hamiltonian term 
\bea
\bm{M}(\bm{r})\cdot\bm{\sigma}&=&M\big[\sin(\bm{q}\cdot\bm{r})\hat{\bm{n}}_{\perp}+\cos(\bm{q}\cdot\bm{r})\hat{\bm{n}}\big]\cdot\bm{\sigma}\no\\
&=&Me^{-i\bm{q}\cdot\bm{r}(\hat{\bm{n}}\times\hat{\bm{n}}_{\perp})\cdot\bm{\sigma}/2}\hat{\bm{n}}\cdot\bm{\sigma}
e^{i\bm{q}\cdot\bm{r}(\hat{\bm{n}}\times\hat{\bm{n}}_{\perp})\cdot\bm{\sigma}/2}\,,\qquad
\eea

\noi while note that the unit vectors $\hat{\bm{n}}$ and $\hat{\bm{n}}_{\perp}$, comprise the complex spin-vector $\hat{\bm{n}}_{\bm{q}}=(\hat{\bm{n}}-i\hat{\bm{n}}_{\perp})/\sqrt{2}$ (up to an overall factor).
For illustration purposes let us momentarily consider free electrons under the influence of the above MH term. Gauging away the spatially varying SU(2) phase above, one obtains the single electron Hamiltonian   
\bea
&&\frac{[\bm{p}-\hbar\bm{q}(\hat{\bm{n}}\times\hat{\bm{n}}_{\perp})\cdot\bm{\sigma}/2]^2}{2m}+M\hat{\bm{n}}\cdot\bm{\sigma}=\frac{(\hat{\bm{q}}_{\perp}\cdot\bm{p})^2}{2m}+\frac{(\hbar\bm{q})^2}{8m}\no\\
&&+\frac{(\hat{\bm{q}}\cdot\bm{p})^2}{2m}-\frac{1}{2}\alpha_{\bm{q}}\hat{\bm{q}}\cdot\bm{p}(\hat{\bm{n}}\times\hat{\bm{n}}_{\perp})\cdot\bm{\sigma}+M\hat{\bm{n}}\cdot\bm{\sigma}\,,
\label{eq:MHInducedSOC}
\eea

\noi where $\alpha_{\bm{q}}=\hbar|\bm{q}|/m$ defines the strength of the generated SOC, while we introduced the unit vector $\hat{\bm{q}}=\bm{q}/|\bm{q}|$. The presence of inversion-breaking SOC mediates a magnetoelectric coupling and can lead to a number of so-called direct and inverse spin galvanic phenomena~\cite{SpinGalvanic}. Here we will focus on the generation of a homogeneous magnetization when an electric charge current ($\bm{I}$) flows through the system or an electric field ($\bm{{\cal E}}$) is applied. Both situations can be described by coupling the electronic system to a spatially homogeneous but generally time-dependent vector potential $\bm{A}(t)$, that in the Coulomb gauge yields the electric field $\bm{{\cal E}}(t)=-\partial_t\bm{A}(t)$. Instead, if $\bm{A}$ is time-independent, it corresponds to a current flow bias, i.e. $\bm{I}\propto\bm{A}$. 

It is straightforward to infer that a vector potential minimally coupled to the momentum, i.e. $\bm{p}\rightarrow \bm{p}+e\bm{A}(t)$, will generate a time-dependent but spatially homogeneous magnetization of the form
\bea
\bm{M}_{\bm{0}}(t)\propto (\hat{\bm{n}}\times\hat{\bm{n}}_{\perp})\hat{\bm{q}}\cdot\bm{A}(t)
\propto (i\hat{\bm{n}}_{\bm{q}}\times\hat{\bm{n}}_{\bm{q}}^*)\hat{\bm{q}}\cdot\bm{A}(t)\,.\quad\label{eq:magnetoelectricity}
\eea

\noi The derivation of the required magnetoelectric coupling leading to the above relation is presented in Appendix~\ref{App:AppendixMagnetoelectricity}.

Interestingly, via magnetization measurements in the zero-external magnetic field limit, one can arrest the magnetoelectrically generated ferromagnetic moment, as long as the current or electric field are orientated along the MH's spatial modulation direction $\hat{\bm{q}}$. Moreover, the spin-orientation of the induced magnetization can provide information regarding the MH's winding plane, since the former is given by $i\hat{\bm{n}}_{\bm{q}}\times\hat{\bm{n}}_{\bm{q}}^*$. To further demystify the nature of such vectors one observes that they can be linked to the magnetic order parameter's winding in coordinate space along the $\bm{q}$ direction, i.e.
\bea
\bm{w}_{\bm{q}}\propto\int_{\rm UC} d\bm{r}\ph \bm{M}(\bm{r})\times\left[\hat{\bm{q}}\cdot\frac{\partial\bm{M}(\bm{r})}{\partial\bm{r}}\right]\propto i\hat{\bm{n}}_{\bm{q}}\times\hat{\bm{n}}_{\bm{q}}^*\,,\quad
\eea

\noi with UC denoting the magnetic unit cell and $\bm{w}_{\bm{q}}$ corresponding to a vector normal to the winding plane defined by the helix.

Based on the above analysis, we find that magnetoelectric phenomena become possible in the IC magnetic phases discussed here when either one of the quantities, $i\hat{\bm{n}}_{1}\times\hat{\bm{n}}_{1}^*$ or $i\hat{\bm{n}}_{2}\times\hat{\bm{n}}_{2}^*$, becomes non-zero. In the former (latter) case the induced SOC involves the electron's momentum along the $\bm{Q}_1$ ($\bm{Q}_2$) direction. For the IC magnetic phases and respective order parameters under investigation, we observe that only the six new phases lead to magnetoelectric phenomena as summarized in Table~\ref{table:ICMagneticSignatures}. 

In fact, the SWC$_{4,2}$ phases consist of two non-coplanar MHs and can give rise to such phenomena for an arbitrary orientation of $\bm{I}$ or $\bm{{\cal E}}$, since both external pro\-ducts are non-zero and non-parallel. In the present case Eq.~\eqref{eq:magnetoelectricity} and Appendix~\ref{App:AppendixMagnetoelectricity} imply that the magnetization retains contributions from both MHs and one thus obtains
\begin{align}
\bm{M}_{\bm{0}}(t)\propto(i\hat{\bm{n}}_{1}\times\hat{\bm{n}}_{1}^*)\bm{Q}_1\cdot\bm{A}(t)+(i\hat{\bm{n}}_{2}\times\hat{\bm{n}}_{2}^*)\bm{Q}_2\cdot\bm{A}(t)\,.
\label{eq:magnetoelectricitySWC}
\end{align}

\noi In contrast, the remaining magnetic phases induce SOC only along one direction, and thus are not capable of exhibiting magnetoelectric phenomena when the external field is not properly aligned. The reason is that three of those consist only of a single MH that essentially singles-out the direction for which such effects become accessible. Interestingly, the same holds for the DPMH phase due to the coplanarity of the underlying MHs. Therefore, we start from Eq.~\eqref{eq:magnetoelectricitySWC} and set $\hat{\bm{n}}_1=\hat{\bm{n}}_2$ that dictates the DPMH phase. The latter simply yields 
\bea
\bm{M}_{\bm{0}}(t)\propto (i\hat{\bm{n}}_{1}\times\hat{\bm{n}}_{1}^*)\left(\bm{Q}_1+\bm{Q}_2\right)\cdot\bm{A}(t)\,,
\eea

\noi revealing that magnetoelectric phe\-no\-me\-na are not ac\-ces\-si\-ble when $\bm{{\cal E}}||(\bm{Q}_1-\bm{Q}_2)$. 

To this end, let us remark that while we focused here on the generation of a ferromagnetic moment when an electric field is applied or a current flow is induced, reciprocal effects are also possible, by virtue of the magnetoelectric coupling (see Appendix~\ref{App:AppendixMagnetoelectricity}). This can be particularly useful for the case of  the itinerant magnets discussed here, which are expected to be good conductors and, thus, screen the externally applied fields. Therefore, depending on the case, it may appear experimentally more feasible to observe a reciprocal magnetoelectric effect, such as inducing a current by an externally-imposed time-dependent homogeneous magnetization or by virtue of the Zeeman coupling to an applied magnetic field. Both need to have an appropriate orientation for the magnetoelectric effects to take place. Based on Appendix~\ref{App:AppendixMagnetoelectricity}, a field of frequency $\omega$, $\bm{M}_0(\omega)$, generates an electric current $\bm{J}(\omega)$ which reads
\bea
\bm{Q}_s\cdot\bm{J}(\omega)\propto (i\hat{\bm{n}}_{s}\times\hat{\bm{n}}_s^*)\cdot\bm{M}_{\bm{0}}(\omega)\,.
\eea
\noi In the above, we projected the current along the directions of the magnetic wavevectors $\bm{Q}_{s}$ ($s=1,2$).

\subsection{Zeeman-field-induced spin-skyrmion crystals}\label{Sec:Skyrmions}

As discussed earlier, the SWC phases are special as, in contrast to the remaining set of magnetic profiles exa\-mi\-ned here, they are characterized by isolated $\pi$-Berry flux sources. This feature is essentially responsible for the observed distinct behavior of these phases regar\-ding the occurrence of magnetoelectric phenomena independent of the orientation of the electric field. When the $\Theta$ symmetry is broken, one obtains $|\bm{M}(\bm{r})|\neq0$ $\forall\bm{r}$ and the SWC phases allow for the engineering of more exotic magnetic textures, the so-called spin-skyrmion crystal phases. The latter are charac\-te\-ri\-zed by a magnetic skyrmion charge retrieved via the Chern number of the magnetic unit vector $\hat{\bm{n}}(\bm{r})=\bm{M}(\bm{r})/|\bm{M}(\bm{r})|$~\cite{Niu}:
\bea
{\cal C}&=&\frac{1}{4\pi}\int_{\rm UC}d\bm{r}\ph\hat{\bm{n}}(\bm{r})
\cdot\left[\frac{\partial\hat{\bm{n}}(\bm{r})}{\partial x}\times\frac{\partial\hat{\bm{n}}(\bm{r})}{\partial y}\right]\no\\
&\equiv&\frac{1}{2\pi}\int_{\rm UC}d\bm{r}\ph\Omega_{xy}(\bm{r})\,,
\eea

\noi with UC denoting the magnetic unit cell, determined by the ordering wavevectors $\bm{Q}_{1,2}$. In fact, the SWC$_4$ phase constitutes a critical magnetic phase, and lies on the border separating two spin-skyrmion crystal phases of dif\-fe\-rent skyrmion charge. 

To engineer a spin-skyrmion crystal from the SWC phases it is required to violate the $\Theta$ symmetry. While it is expected that the incommensurability will violate it in the actual material, the outcome is difficult to predict and, thus, a control knob to induce a non-zero skyrmion charge is certainly de\-si\-ra\-ble. One way to achieve the latter is to externally apply a homogeneous Zeeman field, that is even under tran\-sla\-tions and odd under ${\cal T}$. Via the application of the Zeeman field we are additionally in a position to control the sign of ${\cal C}$~\cite{Mendler}, which also speci\-fies the orientation of the out-of-plane orbital angular momentum in the system~\cite{Volovik}. Note that while an arbitrarily weak Zeeman field violates the $\Theta$ symmetry, ${\cal C}$ is non-zero only within a particular window of the Zeeman energy va\-lues, $E_{\rm Zeeman}$.  

One finds that, given the expressions of Eqs.~\eqref{eq:MW4} and~\eqref{eq:MW2}, the SWC phases become skyrmionic by applying a Zeeman field $\bm{{\cal B}}={\cal B}\hat{\bm{z}}$. The straightforward calculation of ${\cal C}$ yields that $|{\cal C}|=1$ for both phases. In the case of the SWC$_4$ phase the skyrmion charge is non-zero for $|E_{\rm Zeeman}|<2\sqrt{2}M|\sin\lambda|$, i.e. even for an infinitesimal applied field. In contrast, we find that a threshold field is required to induce a non-zero skyrmion charge for the SWC$_2$ phase, which is proportional to the degree of ne\-ma\-ti\-ci\-ty of the magnetic order pa\-ra\-me\-ter. Specifically, the magnetic profile is skyrmionic for $\sqrt{2}M|A-B|<|E_{\rm Zeeman}|<\sqrt{2}M(A+B)$, with $A=\cos\eta$ and $B=\sqrt{2}\sin\eta|\cos\lambda|$. Finally, note that the pre\-sen\-ce of SOC will pin the magnetic moment direction, assumed free in Eqs.~\eqref{eq:MW4} and~\eqref{eq:MW2}, and thus constrain the actual orientation of the required Zeeman field. See also Sec.~\ref{Sec:SOC}.

\section{Incommensurate magnetic phases in the iron-based superconductors}\label{sec:pnictide_section}
 
Having obtained the possible magnetic leading instabilities and their distinctive characteristics for the ge\-ne\-ral class of tetragonal itinerant systems described phenomenologically via the Landau free energy of Eq.~\eqref{eq:free_energy}, we may now proceed by exploring the possibility of such phases appearing in FeSCs. To accomplish this task we extract the coefficients of the free energy functional from two representative multi-orbital microscopic mo\-dels. These are supplemented with standard Hubbard-Hund interactions that can drive the system to a magnetic instability. Below we present the details of this analysis.

\subsection{Microscopic model}\label{Sec:FeSCs}

We adopt a microscopic Hamiltonian, $\mathcal{H}=\mathcal{H}_0 + \mathcal{H}_{\rm int}$, with 
\bea
\mathcal{H}_0 &=& \sum_{\bm{k}} \sum_{\substack{ab \\ \sigma}}\left(\epsilon_{ab}(\bm{k}) - \mu \delta_{ab} \right)c^{\dagger}_{\bm{k} a \sigma} c^{\phantom{\dagger}}_{\bm{k}b\sigma},
\eea

\noi provided by tight-binding fits to DFT calculations. Here we consider two five-orbital bandstructures, adopted from Refs.~\cite{graser10,ikeda10}, appropriate for BaFe$_2$As$_2$ and LaFeAsO, respectively. Although, the BaFe$_2$As$_2$ bandstructure is three-dimensional, here we restrict to the $k_z=0$ plane. The bandstructures are supplemented by $\mathcal{H}_{\rm int}$ consisting of Hubbard-Hund interactions~\cite{Hubbardmodel1,Hubbardmodel2}:
\bea
\mathcal{H}_{\rm int} &=& \frac{U}{\mathcal{N}}\sum_{\bm{q}}\sum_{a}n_{\bm{q}a\uparrow}n_{\bm{-q}a\downarrow} + \frac{U'}{\mathcal{N}}\sum_{\bm{q}}\sum_{\substack{a < b \\ \sigma\sigma'}}n_{\bm{q}a\sigma}n_{\bm{-q}b\sigma'} \nonumber \\ && + \frac{J}{2\mathcal{N}}\sum_{\bm{k}\bm{k}^{\prime}\bm{q}}\sum_{\substack{a \neq b \\ \sigma\sigma'}}c_{\bm{k+q}a\sigma}^{\dagger}c_{\bm{k}b\sigma}c_{\bm{k'-q}b\sigma'}^{\dagger}c_{\bm{k'}a\sigma'} \nonumber \\ && + \frac{J'}{2\mathcal{N}}\sum_{\bm{k}\bm{k}^{\prime}\bm{q}}\sum_{\substack{a\neq b \\ \sigma}}c_{\bm{k+q}a\sigma}^{\dagger}c_{\bm{k'-q}a\bar{\sigma}}^{\dagger}c_{\bm{k'}b\bar{\sigma}}c_{\bm{k}b\sigma}\,,\label{eq:multi_orbital_interaction}
\eea

\noi where, by imposing SO(3) spin-rotational invariance on the sum of the interaction terms and further assuming orbital-independent interactions, we obtain $U'=U-2J$ and $J=J'$. In the above expressions, $c^{\dagger}_{\bm{k}a\sigma}$ ($c_{\bm{k}a\sigma}$) creates (annihilates) an electron in orbital $a$ with momentum $\bm{k}$ and spin $\sigma$. $\mathcal{N}$ denotes the number of momentum states. The above interaction terms are treated at the mean-field level via a Hubbard-Stratonovich decoupling in the magnetic channel, yielding an effective action for the magnetic order parameters. Expanding to quartic order yields an extension of the magnetic free energy in Eq.~(\ref{eq:free_energy}) to include the orbital content of the magnetic order parameters
\begin{widetext}
\bea
	F &=& \sum_{\bm{q}}\sum_{abcd}\big[ (U^{-1})^{abcd} - \chi^{abcd}_0(\bm{q}) \big]\bm{M}^{ab}(\bm{q})\bm{M}^{cd}(-\bm{q}) 
	 + \sum_{\substack{abcd \\ efgh}} \tilde{\beta}^{abcdefgh}_1 \left( \bm{M}_1^{ab} \cdot \bm{M}_1^{cd\ast} \right) \left( \bm{M}_1^{ef}\cdot \bm{M}_1^{gh\ast} \right) \nonumber \\
	&& +  \sum_{\substack{abcd \\ efgh}} \tilde{\beta}^{abcdefgh}_2 \left( \bm{M}_2^{ab} \cdot \bm{M}_2^{cd\ast} \right) \left( \bm{M}_2^{ef}\cdot \bm{M}_2^{gh\ast} \right) 
	+ \frac{1}{2}\sum_{\substack{abcd \\ efgh}}\left(\beta_1 - \tilde{\beta}_1 \right)^{abcdefgh} \left( \bm{M}_1^{ab} \cdot \bm{M}_1^{cd } \right) \left( \bm{M}_1^{ef\ast}\cdot \bm{M}_1^{gh\ast} \right) \nonumber \\
	&& + \frac{1}{2}\sum_{\substack{abcd \\ efgh}}\left(\beta_2 - \tilde{\beta}_2 \right)^{abcdefgh} \left( \bm{M}_2^{ab} \cdot \bm{M}_2^{cd } \right) \left( \bm{M}_2^{ef\ast}\cdot \bm{M}_2^{gh\ast} \right) 
	+ \sum_{\substack{abcd \\ efgh}} g^{abcdefgh} \left( \bm{M}_1^{ab} \cdot \bm{M}_1^{cd\ast} \right) \left( \bm{M}_2^{ef}\cdot \bm{M}_2^{gh\ast} \right) \nonumber \\
	&& + \frac{1}{2}\sum_{\substack{abcd \\ efgh}} \tilde{g}_1^{abcdefgh} \left( \bm{M}_1^{ab} \cdot \bm{M}_2^{cd} \right) \left( \bm{M}_1^{ef\ast}\cdot \bm{M}_2^{gh\ast} \right) + \frac{1}{2}\sum_{\substack{abcd \\ efgh}} \tilde{g}_2^{abcdefgh} \left( \bm{M}_1^{ab} \cdot \bm{M}_2^{cd\ast} \right) \left( \bm{M}_1^{ef\ast}\cdot \bm{M}_2^{gh} \right)\,.
	\label{eq:free_energy_orbitals}
\eea
\end{widetext}

\noi We note that the coefficients of the free energy are tensors in orbital space and are entirely determined by the microscopic bandstructure. The expressions for these coefficients are given in Appendix~\ref{App:AppendixSusceptibility}. Here we retained the $\bm{q}$-dependence of the magnetic order at quadratic level to identify the transition from commensurate to IC magnetism. By contrast, the quartic order terms are calculated for fixed $\bm{q}$, i.e. $\bm{Q}_{1,2}$ which are set by the quadratic term. 

To determine the magnetic transition temperature we follow the procedure described in Ref.~\onlinecite{christensen17} and consider the generalized eigenvalue problem
\bea
\left[{\cal D}^{-1}_{\rm mag}(\bm{q}) \right]^{abcd} v_{cd}(\bm{q}) = \lambda(\bm{q}) v_{ab}(\bm{q})\,,
\eea

\noi where
\bea
\left[{\cal D}_{\rm mag}^{-1}(\bm{q})\right]^{abcd}= (U^{-1})^{abcd} - \chi^{abcd}_0(\bm{q})\,,\label{eq:prop_eq}
\eea

\begin{figure*}[t!]\centering
\includegraphics[width=\textwidth]{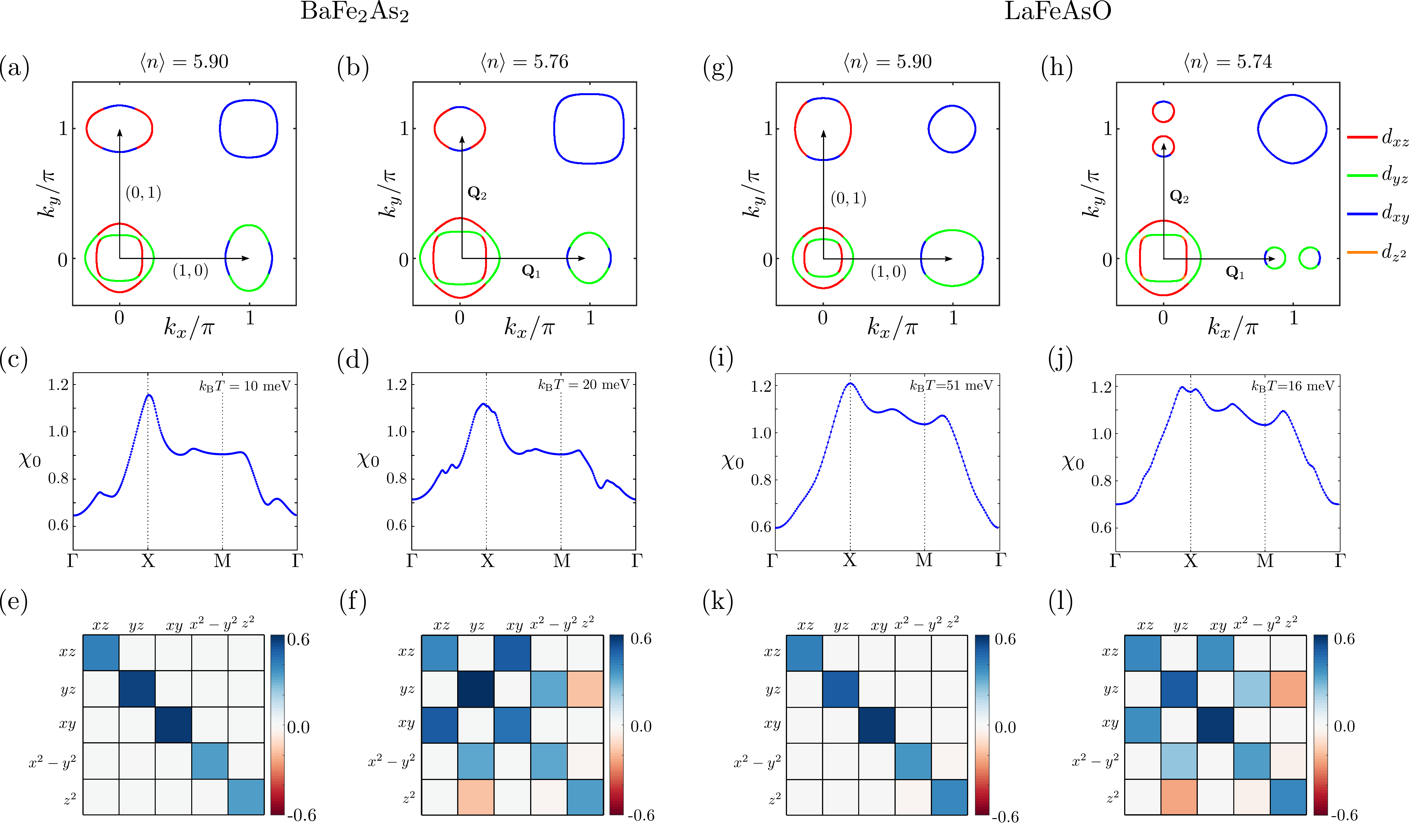}
\caption{Fermi surfaces, bare physical susceptibilities $\sum_{ab}\chi^{aabb}_0(\bm{q})$ and orbital content at the magnetic transitions for (a)--(f) BaFe$_2$As$_2$ and (g)--(l) LaFeAsO. The Fermi surface depicted in (a) for $\langle n \rangle = 5.90$ results in a commensurate magnetic order while the Fermi surface in (b) with $\langle n \rangle = 5.76$ leads to incommensurate magnetic order. We note that both these refer to the $k_z=0$ slice of the BaFe$_2$As$_2$ bandstructure of Ref.~\onlinecite{graser10}. In the commensurate case, the susceptibility is depicted in (c), clearly exhibiting a peak at the X point $(\pi,0)$. In contrast, in the incommensurate case (d), the peak is displaced from $(\pi,0)$ to $(\pi-\delta,0)$. For the case shown in (d), $\langle n \rangle = 5.76$ and $\delta=\pi/50$. Incommensurability in the BaFe$_2$As$_2$ bandstructure onsets at $\langle n \rangle = 5.80$ for which $\delta=\pi/200$. Furthermore, we note that the path between $\Gamma$ and X is dominated by a single peak, and while subleading peaks are present, these are clearly much less pronounced. This allows us to neglect higher harmonics of $\bm{Q}=(Q,0)/(0,Q)$. The orbital content of the magnetic order parameter at the transition is shown in (e) and (f) for the commensurate and incommensurate cases, respectively. In the commensurate case the contributions are purely real. The additional terms appearing as magnetism becomes incommensurate are imaginary, and for purposes of the presentation have been multiplied by a factor of $10$. Hence, in the IC case $\bm{M}(\bm{Q}_{1,2}) \neq \bm{M}(-\bm{Q}_{1,2})$, as this would result in a change of sign of the imaginary parts. In (g) we show the Fermi surface for the LaFeAsO bandstructure with $\langle n \rangle = 5.90$. The commensurate-to-incommensurate transition is brought on by a change in Fermi surface topology, as seen in (h). As in the case of BaFe$_2$As$_2$, the bare susceptibility depicted in (j) shows a clear peak at $(\pi-\delta,0)$, although in this case $\delta\approx\pi/10$ even close to the transition to incommensurate magnetism. The orbital content of the magnetic order parameter depicted in (k) and (l) is very similar to the BaFe$_2$As$_2$ bandstructure, although the $xy$ orbital is more dominant in the case of LaFeAsO.}
\label{fig:Susc_FS_Orb_W}
\end{figure*}

\begin{figure*}[t!]
\centering
\includegraphics[width=\textwidth]{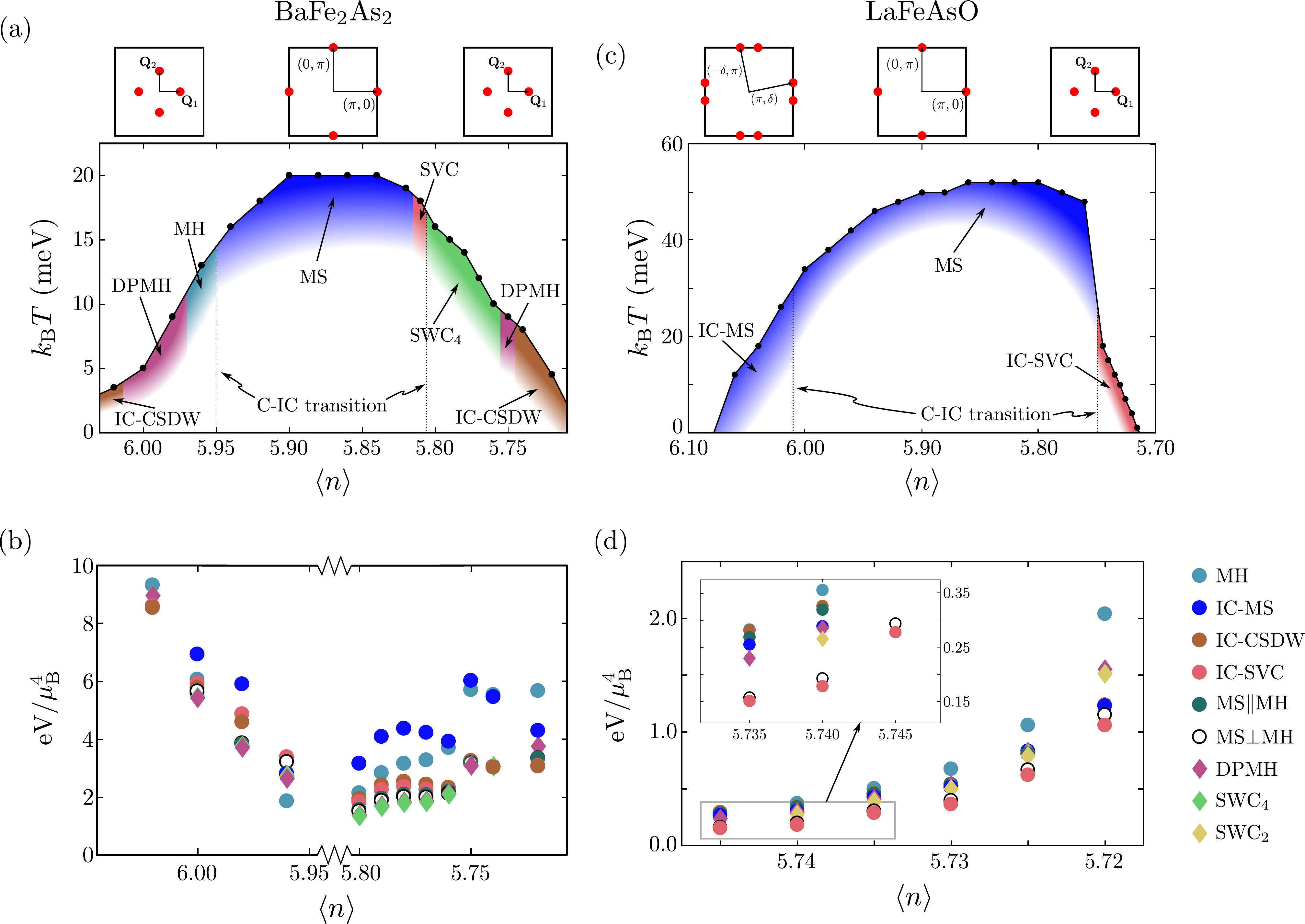}
\caption{Magnetic transition temperatures, leading instabilitites and free energies as a function of the filling $\langle n \rangle$ for two representative five-orbital models corresponding to (a)--(b) BaFe$_2$As$_2$ and (c)--(d) LaFeAsO. The type of incommensurability, $(\pi-\delta,0)/(0,\pi-\delta)$ or $(\pi,\delta)/(-\delta,\pi)$, is illustrated above figures (a) and (c). For BaFe$_2$As$_2$ we find the usual MS phase in the majority of the commensurate region of the phase diagram, see (a). Prior to the transition to an incommensurate phase, a $C_4$ SVC phase sets in. Incommensurate magnetism occurs for $\langle n \rangle > 5.95$ and $\langle n \rangle < 5.81$. In both cases $\bm{Q}_{1,2}=(\pi-\delta,0)/(0,\pi-\delta)$, with $\delta$ evolving smoothly from zero. We find the MH, DPMH, and IC-CSDW phases for $\langle n \rangle >5.95$, and the SWC$_4$, DPMH and IC-CSDW phases for $\langle n \rangle < 5.81$. In (b) the free energies of the various phases in the IC region are shown, indicating the relative proximity of all the IC phases. The full expressions are provided in Appendix~\ref{App:MagPhases}. In (c), the leading instabilities for LaFeAsO are depicted. The electron doped side with $\langle n \rangle > 6$ exhibits IC magnetic order with ordering vector $\bm{Q}_{1,2}=(\pi,\delta)/(-\delta,\pi)$ and IC-MS order. In the commensurate region, the usual MS phase is found. Magnetism becomes incommensurate again on the hole-doped side for $\langle n \rangle < 5.75$, although in this case $\bm{Q}_{1,2}=(\pi-\delta,0)/(0,\pi-\delta)$. However, this commensurate-to-incommensurate transition occurs due to the change in topology of the Fermi surface shown in Fig.~\ref{fig:Susc_FS_Orb_W}(h) and $\delta$ jumps from zero to a finite value, $\delta\approx\pi/10$. The magnetic phase in the IC region is the IC-SVC phase. (d) Magnitude of $4F^{(4)}/M^4$ for the different IC magnetic phases for $U=0.95$eV and $J=U/4$. As the magnetic order becomes incommensurate, the IC-SVC phase is favored. The inset shows a zoom of the lower left corner.}
\label{fig:phase_diagrams}
\end{figure*}

\noi denotes the (inverse) magnetic propagator, $\lambda(\bm{q})$ is the smallest eigenvalue and $v_{ab}(\bm{q})$ the associated eigenmatrix. The vanishing of $\lambda(\bm{Q})$ signals the onset of magnetic order with ordering vector $\bm{Q}$. From the vanishing of $\lambda(\bm{Q})$ we find three types of ordering vectors appea\-ring as a function of electron filling (see insets in Fig.~\ref{fig:phase_diagrams}). The orbital weight of the order parameter with or\-de\-ring vector $\bm{Q}$ is contained in the matrix $v_{ab}(\bm{Q})$ and we write $\bm{M}_{1,2}^{ab}=\bm{M}_{1,2}v_{ab}(\bm{Q}_{1,2})$. The quadratic coefficient provides information on the transition temperature, and the quartic coefficients are required for the determination of the leading magnetic instability. The relation between the quartic coefficients and the microscopic bandstructure is provided by the above Hubbard-Stratonovich decoupling, and due to the IC nature of the ordering vectors it is necessary to truncate the expressions. We accomplish this by assuming that only the lowest har\-mo\-nics contribute, an assumption justified by comparing the magnitude of the peak at $\bm{Q}_{1,2}$ in the bare suscepti\-bi\-li\-ty with the peaks at higher integer multiples of $\bm{Q}_{1,2}$, as shown in Figs.~\ref{fig:Susc_FS_Orb_W}(d) and (j), for the BaFe$_2$As$_2$  and LaFeAsO bandstructures. The eigenmatrix $v_{ab}(\bm{q}=\bm{Q}_{1,2})$ encodes the orbital content of the magnetic order parameter at the magnetic transition~\cite{christensen17}. Note that this object contains information solely about the magnetic order parameter's orbital structure, and not its magnitude. Below, we employ the matrices $v_{ab}(\bm{Q}_{1,2})$ to project Eq.~\eqref{eq:free_energy_orbitals} onto the leading instability. Thus, we obtain the free energy of Eq.~\eqref{eq:free_energy} with the coefficients obtained from the microscopic models, see Eqs.~\eqref{eq:betatilde1}--\eqref{eq:gtilde2} of Appendix~\ref{App:AppendixSusceptibility}.

The incommensurability evident in Figs.~\ref{fig:Susc_FS_Orb_W}(d) and (j) is caused by a change in the Fermi surface nesting pro\-per\-ties induced by sufficient hole or electron doping. In Figs.~\ref{fig:Susc_FS_Orb_W}(a) and (g) we show the Fermi surfaces for both the BaFe$_2$As$_2$ and LaFeAsO bandstructures in the paramagnetic phase, at a doping leading to commensurate magnetic order, i.e. $\langle n \rangle =5.90$. We also show the Fermi surfaces for systems with substantial hole doping, leading to IC magnetic orders, see Figs.~\ref{fig:Susc_FS_Orb_W}(b) and (h). Note that the doping necessary to drive the IC transition is substantial, i.e. the commensurate order is preserved even under mo\-de\-ra\-te doping, consistent with experimental observations. In the case of BaFe$_2$As$_2$ the Fermi surface is not drastically modified as the magnetic transition becomes incommensurate, see Figs.~\ref{fig:Susc_FS_Orb_W}(a) and (b). For LaFeAsO, on the other hand, the IC transition occurs as a result of the change in topology of the electron pockets at X and Y, see Fig.~\ref{fig:Susc_FS_Orb_W}(h). This is reflected in the suppression of the magnetic transition temperature in Fig.~\ref{fig:phase_diagrams}(c). In Fig.~\ref{fig:Susc_FS_Orb_W} we also present the orbital content obtained from $v_{ab}(\bm{Q}_{1,2})$ for both commensurate [(Figs.~\ref{fig:Susc_FS_Orb_W}(e) and (k)] and IC magnetism [Figs.~\ref{fig:Susc_FS_Orb_W}(f) and (l)] for the two bandstructures. We observe, that, as the ordering vector becomes IC, the orbitally resolved magnetic order parameter acquires imaginary components, an indication that indeed $\bm{M}(\bm{Q}_{1,2}) \neq \bm{M}(-\bm{Q}_{1,2})$, as expected.

\subsection{Magnetic phase diagram}

We determine the location of the magnetic transition and the respective leading instability as a function of the electron filling $\langle n \rangle$, for $U=0.95$~eV and $J=U/4$, for the two bandstructures described above. The results are depicted in Fig.~\ref{fig:phase_diagrams}. In both cases we find a large region of commensurate magnetism with $\bm{Q}=(\pi,0)/(0,\pi)$, around a filling of $\langle n \rangle \approx 5.90$ with IC phases appearing upon either hole- or electron-doping, as the Fermi surface nesting properties are sufficiently modified. As expected based on the picture of iti\-ne\-rant magnetism, the magnetic order becomes IC as the Fermi surface is substantially deformed by the addition or removal of carriers. For the BaFe$_2$As$_2$ bandstructure, both electron- and hole-doping leads to IC phases with $\bm{Q}_{1,2}=(\pi-\delta,0)/(0,\pi-\delta)$, with $\delta$ increasing smoothly from zero. The transition temperature broadly follows the expected behavior~\cite{Inflection,chubukov10}, exhibiting an inflection point at the commensurate-to-incommensurate transition, see Fig.~\ref{fig:phase_diagrams}(a). On the electron-doped side, we find two C$_2$-symmetric phases, in addition to the C$_4$-symmetric IC-CSDW phase. These are the MH and DPMH phases. On the hole-doped side we find a transition to a C$_4$ SVC phase prior to the commensurate-to-incommensurate transition. As magnetism becomes incommensurate, the SWC$_4$ phase becomes favored, followed by the DPMH and the IC-CSDW phases.

In contrast, for the LaFeAsO bandstructure, electron-doping yields a region with $\bm{Q}_{1,2}=(\pi,\delta)/(-\delta,\pi)$ and IC magnetic stripe order, as seen in Fig.~\ref{fig:phase_diagrams}(c). In this case $\delta$ also smoothly increases from zero. Substantial hole-doping (i.e. $\langle n \rangle \approx 5.75$) yields the IC magnetic wavevectors $\bm{Q}_{1,2}=(\pi-\delta,0)/(0,\pi-\delta)$, similar to the case of the BaFe$_2$As$_2$ bandstructure. However, in this case $\delta$ exhibits a jump from zero to a finite value, associated with the change in topology of the Fermi surface shown in Fig.~\ref{fig:Susc_FS_Orb_W}(h). This also leads to a suppression of the magnetic transition temperature, as seen in Fig.~\ref{fig:phase_diagrams}(c). The commensurate magnetic stripe phase occupying a large part of the phase diagram is succeeded by the IC-SVC phase, which persists until the end of the magnetic dome.

In Fig.~\ref{fig:phase_diagrams}(b) and (d) we show the free energy of the va\-rious phases in the IC region. More specifically, we plot the value of $4F^{(4)}/M^4$, where $F^{(4)}$ is the quartic part of the free energy in Eq.~\eqref{eq:free_energy}, as discussed in Appendix~\ref{App:MagPhases}. Note that this number does not take into account the effect of the increasing magnitude of the magnetic order parameter. Thus, it is only valid in the vicinity of the transition. For the BaFe$_2$As$_2$ bandstructure, many phases appear in close proximity, leading to a number of different leading instabilities, as seen in Fig.~\ref{fig:phase_diagrams}(a). In the case of LaFeAsO [Fig.~\ref{fig:phase_diagrams}(d)], the IC-SVC phase is favored throughout the IC region, although the remaining phases appear close in energy. 

\begin{figure}[t!]
\centering
\includegraphics[width=1.0\columnwidth]{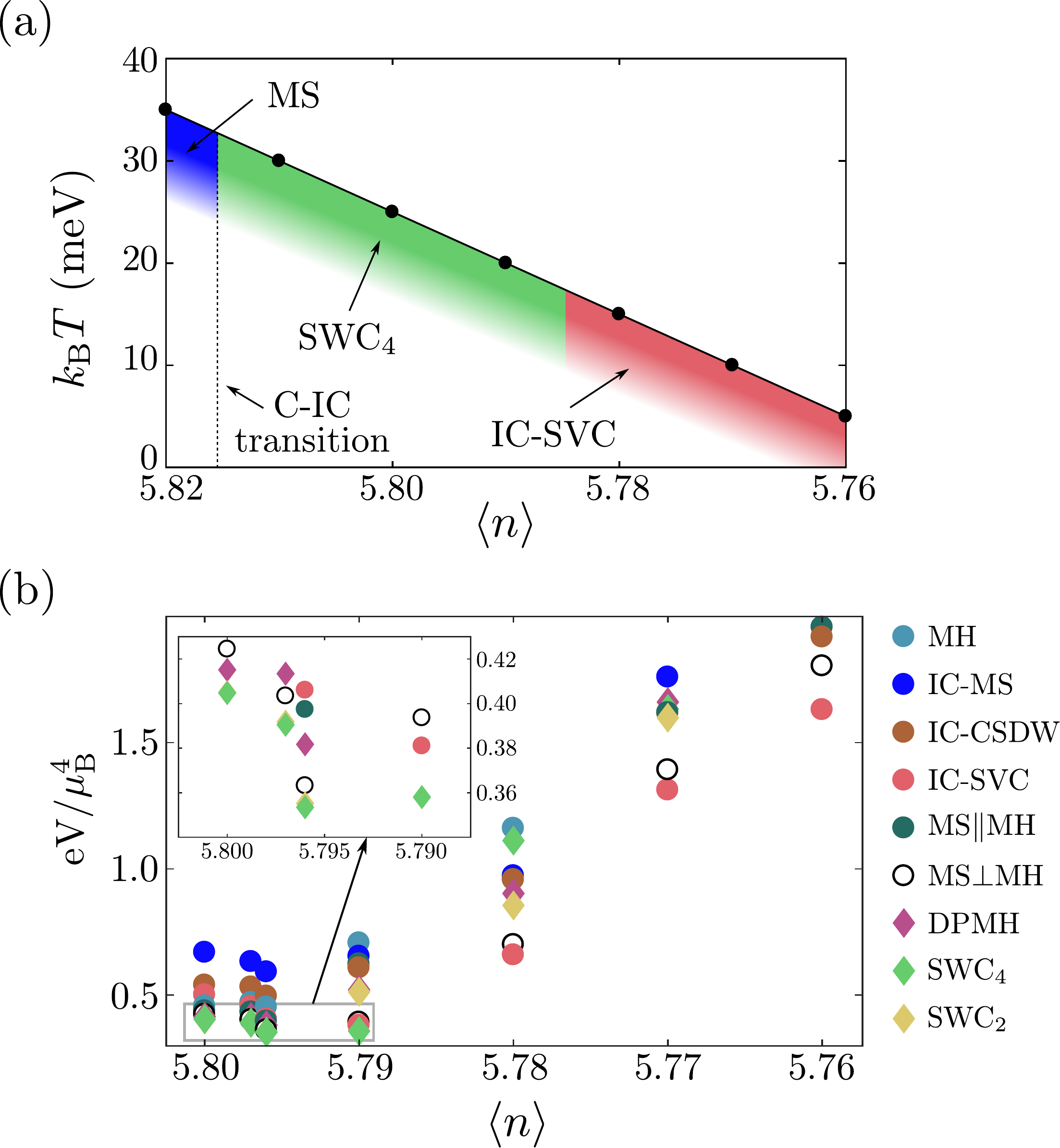}
\caption{\label{fig:orb_int_energies}(a) Phase diagram for the LaFeAsO bandstructure of Ref.~\onlinecite{ikeda10} with $U=0.86$ eV, $J=U/4$ and $U_{xy}=1.1U$, i.e. the $xy$ orbital is subject to a stronger interaction. The slight change in interaction parameters has changed the ground state in a region of the phase diagram, from to IC-SVC to SWC$_4$. The degree of incommensurability, $\delta$, is the same as the one shown in Fig.~\ref{fig:phase_diagrams}(c). (b) Value of $4F^{(4)}/M^4$ for the case with orbital-dependent interactions. The SWC$_4$ phase is favored close to $\langle n \rangle \approx 5.80$.}
\end{figure}

The small free-energy separations observed in Figs.~\ref{fig:phase_diagrams}(b) and (d) suggest that the phases stabilized are sensitive to specific details of the microscopic mo\-dels. To highlight this sensitivity, we consider the effect of small changes to the interaction parameters, inspired by recent proposals of orbital selectivity~\cite{Orbsel1,Orbsel2,Orbsel3,OrbitalSelectivityScience, OrbitalSelectivityPRB}. Hence, we allow the interactions in Eq.~\eqref{eq:multi_orbital_interaction} to exhibit orbital dependence. Despite the fact that such an orbital dependence is ty\-pi\-cal\-ly ignored, it is well-known to be present in rea\-li\-stic mo\-dels, and recent experimental and theoretical works~\cite{Orbsel1,Orbsel2,Orbsel3,OrbitalSelectivityScience, OrbitalSelectivityPRB} have re-emphasized its important role. Nevertheless, the consideration of such an orbital dependence introduces a large pa\-ra\-me\-ter space and an exhaustive treatment is beyond the scope of the current work. Instead, we focus on varying the interactions of the $xy$ orbital. This is expected to be the most relevant for the present analysis since it is associated with the largest orbital weight as seen in Fig.~\ref{fig:Susc_FS_Orb_W}. To assess the impact of such orbital dependence we vary the $xy$ component of $(U^{-1})^{abcd}$ in Eq.~(\ref{eq:prop_eq}) and repeat the above analysis for the LaFeAsO band. We take $U=0.86$eV, $U_{xy}=1.1 U \approx 0.95$eV and $J=U/4$. Unsurprisingly, we find a magnetic dome with a slightly reduced transition temperature compared to the one shown in Fig.~\ref{fig:phase_diagrams}(c). A commensurate-to-IC transition occurs around $\langle n \rangle \approx 5.805$ and, inte\-re\-sting\-ly, the lowest energy phase in the imme\-dia\-te vicinity of this transition is the SWC$_4$, as seen in Fig.~\ref{fig:orb_int_energies}(a).

Based on Figs.~\ref{fig:phase_diagrams} and \ref{fig:orb_int_energies}, we verify that the magnetic phase diagram is sensitive to both the bandstructure and the interaction parameters. The appearance of the SWC$_4$ phase implies that the FeSCs are indeed potential candidates for textured magnetic phases and, given the reported coexistence of magnetism and super\-con\-duc\-ti\-vity~\cite{johrendt11,avci11,klauss15,avci14a,wang16a,ni08a,nandi10a}, opens perspectives for realizing intrinsic topological superconductivity~\cite{steffensen}.

\subsection{Effects of spin-orbit coupling}\label{Sec:SOC}

So far we have investigated the accessible magnetic phases via the free energy of Eq.~\eqref{eq:free_energy}, which neglects the effects of SOC. Here we extend our approach in order to include the effects of a SOC respecting the symmetries of the system~\cite{cvetkovic13}. At leading order, the free energy is modified by the addition of the term
\bea
\delta F&=&\alpha_{1}\left(|M_{1,x}|^2+|M_{2,y}|^2\right)+\alpha_2\left(|M_{2,x}|^2+|M_{1,y}|^2\right)\no\\ 
&+&\alpha_3 \left(|M_{1,z}|^2+|M_{2,z}|^2\right),\label{eq:free_energy_SOC}
\eea

\noi where the $\alpha$-coefficients can be determined from a microscopic model including an atomic $\bm{L}\cdot\bm{S}$ coupling~\cite{christensen15,scherer17}. Note, however, that a complete picture regarding SOC in FeSCs requires the consideration of a ten band model~\cite{TenBand1,TenBand2}, taking into account the two inequivalent Fe sites of the FeAs layers. Lastly, note that experimental evidence points to $\alpha_1$ being the smallest for underdoped compounds, as these display in-plane magnetic order. For systems approaching optimal hole-doping, the magnetic moments reorient out-of-plane~\cite{allred16a,wasser15}, indi\-ca\-ting that $\alpha_3$ is the smallest~\cite{luo2013,zhang,qureshi2014,song16}.

As the above term adds a quadratic contribution to the free energy, it plays an important role in selecting the magnetic order at the leading instability. Moreover, the additional terms modify the magnetic transition temperature $T_{\rm mag}$, obtained for vanishing SOC. Hence $T_{\rm mag}$ is now replaced by $T_{\rm mag}^{\rm SOC}$. The normalized difference $\delta T_{\rm mag}=(T_{\rm mag}^{\rm SOC}-T_{\rm mag})/T_{\rm mag}$ can be either positive or negative and yields a quantitative measure of the SOC strength. Here we assume that the SOC is weak~\cite{NoteSOC1}, i.e. $|\delta T_{\rm mag}|\ll1$, a realistic assum\-ption for FeSCs~\cite{Borisenko}. The presence of SOC also implies that only certain components of the magnetic order parameter can condense below the critical temperature $T_{\rm mag}^{\rm SOC}$, and these are selected by the smallest of the $\alpha$-coefficients. At first sight this would imply that the energetics shown in Figs.~\ref{fig:phase_diagrams} and~\ref{fig:orb_int_energies}, obtained without SOC, can be drastically altered when $\delta F$ is added. However, as we discuss below, the results obtained with SOC remain relevant by virtue of secondary transitions. 

As seen from Eq.~(\ref{eq:free_energy_SOC}), if either $\alpha_1$ or $\alpha_2$ is the smal\-lest the magnetic moments align in the FeAs plane, either parallel or perpendicular to the ordering vector. If instead $\alpha_3$ is the smallest, the moments point out-of-plane. Thus, only a subset of the nine IC magnetic phases found earlier are compatible with the magnetic moment directions fixed by the SOC. This subset consists of the IC generalizations of the three commensurate phases: IC-\{MS,\,CSDW,\,SVC\}. The possible emergence of the IC-MS is completely unaffected by the presence of SOC, as can be inferred from Eq.~(\ref{eq:free_energy_SOC}). On the other hand, the appearance of the IC-SVC (IC-CSDW) is hindered if $\alpha_{3}$ ($\alpha_{1}$ or $\alpha_2$) is the smallest coefficient. In such si\-tua\-tions the IC-SVC or IC-CDSW will only appear as the result of secondary transitions. Similarly, this is always the case for the six textured magnetic phases which are ge\-ne\-ral\-ly disfavored when SOC is present. The appea\-rance of these phases becomes possible only if they minimize the quartic term $F^{(4)}$ of the free energy. Speci\-fi\-cal\-ly, these transitions can occur once $T$ is lowered sufficiently to allow for the modulus of $M$ to grow, and the quartic part $F^{(4)}$ to overcome the SOC energy. At the temperature $T_{\rm mag}^{(2)}$, where this secondary transition takes place, the magnetic order reorganizes and develops additional components in order to minimize $F^{(4)}$~\cite{NoteSOC2}.

We proceed by providing the implications of a non-negligible SOC on the phase diagrams depicted in Figs.~\ref{fig:phase_diagrams}(a) and (c) and in Fig.~\ref{fig:orb_int_energies}(a). The IC-MS region is unaffected by the addition of SOC. On the other hand, the IC-CSDW and IC-SVC are affected if $\alpha_{1,2}$ and $\alpha_3$ are the smallest, respectively. Starting with BaFe$_2$As$_2$ and $\alpha_{1}$ (or $\alpha_2$) being the smallest, we find that the free energy hierarchy of Fig.~\ref{fig:phase_diagrams}(b) implies that the IC phases of Fig.~\ref{fig:phase_diagrams}(a), DPMH, IC-CSDW, and SWC$_4$ are all replaced by the IC-SVC phase, while the MH is replaced by the IC-MS phase. On the contrary, if $\alpha_3$ is the smal\-lest, the IC-CSDW phase replaces the other IC phases, while the commensurate SVC phase is replaced by the commensurate CSDW phase. The phase diagram in the case of LaFeAsO, Fig.~\ref{fig:phase_diagrams}(c), is unchanged if $\alpha_{1}$ (or $\alpha_2$) is the smallest. On the other hand, if $\alpha_3$ is the smal\-lest, the IC-MS phase is stabilized in the IC regions of the phase diagram, as can be gleaned from Fig.~\ref{fig:phase_diagrams}(d). In Fig.~\ref{fig:orb_int_energies} the SWC$_4$ phase is replaced by the IC-SVC phase in the case where $\alpha_1$ (or $\alpha_2$) is the smallest. In the case where $\alpha_3$ is the smal\-lest, the IC-CSDW phase replaces both the SWC$_4$ and the IC-SVC phases. Note that, for $\langle n \rangle = 5.77$ the IC-CSDW phase is nearly degenerate with the MS$\parallel$MH phase, and thus hidden in Fig.~\ref{fig:orb_int_energies}.

Note that, whilst the phases to be replaced do not constitute leading instabilities in the presence of SOC, they still minimize the $F^{(4)}$ part of the free energy. As a result, additional magnetic order parameter components will appear at lower tempe\-ra\-tu\-res, possibly resulting in an admixture of phases~\cite{MoSOC}. In this manner, all the phases predicted by the spin-isotropic free ener\-gy are recovered through secondary phase transitions, provided we remain within the regime of weak SOC. We remark that the above analysis holds when one of the $\alpha_{1,2,3}$ coefficients is smaller than the other two. Otherwise, possible degeneracies between pairs of these coefficients can allow for textured phases to survive as leading instabilities or significantly narrow down the difference $|T_{\rm mag}^{\rm SOC}-T_{\rm mag}^{(2)}|$. Interestingly, the smaller this difference becomes the more difficult it may be to experimentally distinguish the two transitions.

\subsection{Connection to the unknown C$_2$-symmetric magnetic phase of Ba$_{1-x}$Na$_x$Fe$_{2}$As$_{2}$}\label{Sec:Unknown}

Here we comment on the connection between our fin\-dings and the experimental results presented in Ref.~\onlinecite{wang16a}. In Ref.~\onlinecite{wang16a}, an additional C$_2$ symmetric phase was observed, which was shown to be distinct from the usual MS phase. It is challenging to account for such a phase within the usual commensurate scenario. Indeed, there are no commensurate candidates resulting from second-order phase transitions from the paramagnetic phase. On the other hand, allowing for IC magnetism introduces six additional phases, five of which are C$_2$-symmetric. As discussed above however, the presence of SOC also precludes the appearance of these phases as a result of a second-order phase transition from the paramagnetic phase. The experimental observations can be reconciled with the above facts provided that the magnetic transition is first-order. In this case, the intermediate IC-MS, IC-SVC or IC-CSDW dictated by the finite SOC is no longer required, and a direct transition between the paramagnetic phase and one of the five new textured IC C$_2$-symmetric phases is possible. At this stage, very little is known about the properties of the newly observed C$_2$ symmetric magnetic phase. Therefore, to resolve the nature of this phase additional experimental results are warranted, in particular to confirm whether the magnetic order is incommensurate in this region or not.

\section{Conclusions and Outlook}\label{Sec:Conclusions}

In the present work we shed light on aspects of IC magnetism and provided a ge\-ne\-ral classification of the possible magnetic phases that appear at the paramagnetic-magnetic transition for particular types of incommensurability. Our study is motivated by the marked tendency of the FeSCs towards magnetism and the observation of a puzzling C$_2$-symmetric phase, distinct from the magnetic stripe, in Na-doped BaFe$_2$As$_2$~\cite{wang16a}. 

By employing two realistic five-orbital models, we demonstrated that such IC scenarios are feasible in these systems. Our findings reveal that a subset of these new phases consisting of the magnetic helix (MH), double parallel magnetic helix (DPMH) and the C$_4$-symmetric spin-whirl crystal (SWC$_4$), can emerge in the BaFe$_2$As$_2$ bandstructure upon electron- or hole-doping. On the other hand, an IC spin-vortex crystal phase (IC-SVC) can emerge on the substantially hole-doped side of the LaFeAsO phase diagram. A C$_4$-symmetric spin-whirl crystal phase is made possible in this case by employing orbital-dependent interactions.

To detect the IC phases studied here in FeSCs and other materials, suitable experimental methods have to be employed. Si\-mi\-lar to the predictions of Ref.~\onlinecite{MariaSTM} for the three standard commensurate magnetic phases, experimental fingerprints of the novel IC phases are also expected to become evident in spin-resolved scanning tunne\-ling microscopy. However, since directly probing the spin-structure of the order pa\-ra\-me\-ter by e.g. polarized neutron scattering can be challenging, we proposed alternative routes for diagnosing the underlying magnetic order. These indirect measurements rely on inferring the induced charge order and magnetoelectric coupling. The induced charge order is expected to be detectable via the observation of Bragg peaks at particular wavevectors from the set $\{\pm2\bm{Q}_{1,2},\pm\bm{Q}_1\pm\bm{Q}_2\}$, depending on the magnetic phase. The magnetic helix, the IC spin-vortex crystal and the double parallel magnetic helix phases can be directly distinguished via this method. In contrast, the remaining six order parameters can be uniquely identified by complementary magnetoelectric measurements. Textured magnetic phases induce inversion-symmetry-breaking SOC that makes direct and inverse spin-galvanic effects accessible. In more detail, one can induce a ferromagnetic moment via a current flow or the application of an electric field, while reciprocal phenomena are also possible. The orien\-tation of the induced magnetization is given by the spin-vectors' cross products $i\hat{\bm{n}}_{s}\times\hat{\bm{n}}_{s}^*$ ($s=1,2$), while its appearance strictly depends on the orientation of the above mentioned external perturbations.   

In addition to these features, the spin-whirl crystal phases are of particular interest, since they can acquire a non-zero skyrmion charge via applying a Zeeman field, but more importantly they can be employed for reali\-zing intrinsic two-dimensional topological superconductors. The possible microscopic coexisten\-ce of textured phases with spin-singlet superconductivity opens the door for actuali\-zing \textit{intrinsic} topological systems harboring Majorana fermions~\cite{KotetesClassi,Nakosai2013,Ojanen,Sedlmayr,Mendler,Chen,Loss}. The latter become accessible by virtue of the inversion-symmetry breaking SOC and magnetoelectric effects induced by the textured phases. Hence, a transition from a topologically trivial to a non-trivial SC phase is expected to take place as the system enters the coexistence phase between IC magnetism and superconductivity. As we demonstrated, textured phases can be stabilized in FeSCs, with the most prominent for engineering topological superconductivity being the C$_4$-symmetric spin-whirl crystal phase, found as a stable minimum of the free energy for both bandstructure types considered here.

Interestingly, the IC nature of these magnetic textures implies that a non-zero ferromagnetic moment will be ge\-ne\-ra\-ted in a finite-sized two-dimensional FeSC. In the event that this net-magnetization field is felt by the entire bulk (only by the surface) of the FeSC material, the resul\-ting magnetic superconductor will belong to the class of chiral (helical) to\-po\-lo\-gi\-cal superconductors. In the case of a gapped bulk energy spectrum, these magnetic superconductors are cha\-racterized by a $\mathbb{Z}$ ($\mathbb{Z}_2$) topo\-lo\-gi\-cal invariant~\cite{QiZhang,KotetesClassi} that, as long as bulk-boundary correspondence remains intact~\cite{MTM}, yields the number of chiral (helical) Majorana modes per edge. Instead, if the bulk energy spectrum exhibits nodal gap closings, then flat band or other more complex types of Majorana edge modes become accessible~\cite{steffensen}.

The potential experimental observation of microscopic coe\-xi\-stence of magnetism and super\-con\-duc\-ti\-vi\-ty in FeSCs opens novel paths for crafting topological superconductors, distinct from the already existing me\-cha\-nisms involving FeSe~\cite{TenBand2,FeTeSeTopo1,Xue,FeTeSeTopo2,shin18} or other hybrid structures consisting of two dimensional magnetic tex\-tures~\cite{Nakosai2013,Sedlmayr,Mendler,Chen,Loss} in pro\-xi\-mi\-ty to conventional superconductors.

\begin{acknowledgments}
The authors gratefully acknowledge D. Steffensen, D. D. Scherer and M. N. Gastiasoro for inspiring and helpful discussions. M.~H.~C. and B.~M.~A. acknow\-ledge financial support from a Lundbeckfond fellowship (Grant No. A9318). P.~K. and B.~M.~A. acknowledge support from the Independent Research Fund Denmark grant number DFF-6108-00096.
\end{acknowledgments}

\appendix

\newpage

\begin{widetext}

\section{Incommensurate Magnetic Phases}\label{App:MagPhases}

As we showed in the main text, there exist nine distinct incommensurate (IC) magnetic phases which extremize the Landau functional considered in Eq.~(1). For each one of these phases we present: the configuration of the respective $\hat{\bm{n}}_{1,2}$ spin-vectors, the corresponding (or other symmetry-equivalent) magnetization profile $\bm{M}(\bm{r})$, and also the corresponding normalized and shifted quartic free energy term ${\cal F}^{(4)}\equiv 4F^{(4)}/M^4-2\tilde{\beta}$ (the quadratic term is the same for all phases). 

\subsubsection{Incommensurate Magnetic Stripe} This single-$\bm{Q}$ C$_2$-symmetric phase has a commensurate analog, appears only for $\beta-\tilde{\beta}<0$, and has the following characteristics:
\bea
\hat{\bm{n}}_1=(0,0,1)\ph{\rm and}\ph\hat{\bm{n}}_2=(0,0,0)\,,\phd\bm{M}(\bm{r})=2M\left(0\,,0\,,\cos(\bm{Q}_1\cdot\bm{r})\right)\phd{\rm and}\phd{\cal F}^{(4)}=2(\beta-\tilde{\beta})\,.
\eea
\subsubsection{Incommensurate Charge-Spin Density Wave} This double-$\bm{Q}$ C$_4$-symmetric phase has a commensurate analog, appears only for $\tilde{g}<0$, and is described by:
\bea
\hat{\bm{n}}_{1}=\left(0\,,0\,,1\right)\ph{\rm and}\ph\hat{\bm{n}}_{2}=\left(0\,,0\,,1\right)\,,\ph 
\bm{M}(\bm{r})=2M\left(0\,,0\,,\cos(\bm{Q}_1\cdot\bm{r})+\cos(\bm{Q}_2\cdot\bm{r})\right)
\ph{\rm and}\ph{\cal F}^{(4)}=2(\beta-\tilde{\beta})\frac{G+\tilde{G}+1}{2}\,.\qquad
\eea
\subsubsection{Incommensurate Spin-Vortex Crystal} This double-$\bm{Q}$ C$_4$-symmetric phase has a commensurate analog, appears only for $\tilde{g}>0$, and has the following form:
\bea
\hat{\bm{n}}_{1}=\left(0\,,0\,,1\right)\ph{\rm and}\ph\hat{\bm{n}}_{2}=\left(0\,,1\,,0\right)\,,\phd
\bm{M}(\bm{r})=2M\left(0\,,\cos(\bm{Q}_2\cdot\bm{r})\,,\cos(\bm{Q}_1\cdot\bm{r})\right)
\phd{\rm and}\phd{\cal F}^{(4)}=2(\beta-\tilde{\beta})\frac{G+1}{2}\,.
\eea

\subsubsection{Magnetic Helix} This single-$\bm{Q}$ C$_2$-symmetric phase is new and appears only for $\beta-\tilde{\beta}>0$. One finds:
\bea
\hat{\bm{n}}_1=\frac{1}{\sqrt{2}}(i,0,1)\ph{\rm and}\ph\hat{\bm{n}}_2=(0,0,0)\,,\phd
\bm{M}(\bm{r})=2M\left(\sin(\bm{Q}_1\cdot\bm{r})\,,0\,,\cos(\bm{Q}_1\cdot\bm{r})\right)\phd{\rm and}\phd{\cal F}^{(4)}=0\,.
\eea
\subsubsection{Incommensurate Magnetic Stripe \& $||$ Magnetic Helix} This double-$\bm{Q}$ C$_2$-symmetric phase is new and we obtain:
\bea
\hat{\bm{n}}_{1}=\left(0\,,0\,,1\right)\phd{\rm and}\phd
\hat{\bm{n}}_{2}=\left(i\sin\lambda\,,0\,,\cos\lambda\right)\,,\phd{\cal F}^{(4)}=2(\beta-\tilde{\beta})\frac{(\tilde{G}+2G)^2}{(\tilde{G}+2)^2+8(G-1)}\no\\
\phd{\rm and}\phd\bm{M}(\bm{r})=2M\left(\sin\eta\sin\lambda\sin(\bm{Q}_2\cdot\bm{r})\,,0\,,\cos\eta\cos(\bm{Q}_1\cdot\bm{r})+\sin\eta\cos\lambda\cos(\bm{Q}_2\cdot\bm{r})\right)\,,
\eea

\noi with:
\bea
\cos(2\eta)=-\frac{\tilde{G}^2-4}{(\tilde{G}+2)^2+8(G-1)}\qquad{\rm and}\qquad\cos(2\lambda)=-\tilde{G}\frac{2G+\tilde{G}}{\tilde{G}(\tilde{G}+2)+4(G-1)}\,.
\eea
\noi Note that the helix becomes isotropic, i.e. $\lambda=\pi/4$, only for $\tilde{G}=0$.

\subsubsection{Incommensurate Magnetic Stripe \& $\perp$ Magnetic Helix} This double-$\bm{Q}$ C$_2$-symmetric phase is new and has the following features:
\bea
\hat{\bm{n}}_{1}=\left(0\,,0\,,1\right)\phd{\rm and}\phd\hat{\bm{n}}_{2}=\frac{1}{\sqrt{2}}\left(i\,,1\,,0\right)\,,\phd
\bm{M}(\bm{r})=\sqrt{2}M\left(\sin\eta\sin(\bm{Q}_2\cdot\bm{r})\,,\sin\eta\cos(\bm{Q}_2\cdot\bm{r})\,,\sqrt{2}\cos\eta\cos(\bm{Q}_1\cdot\bm{r})\right)\no\\
\ph{\rm and}\ph{\cal F}^{(4)}=2(\beta-\tilde{\beta})\frac{G^2}{2G-1}\ph{\rm with}\ph\cos(2\eta)=\frac{1}{2G-1}\,.\qquad\qquad\qquad\qquad\qquad\qquad\quad
\eea
\subsubsection{Douple Parallel Magnetic Helix} This double-$\bm{Q}$ C$_2$-symmetric phase (in spite of $\eta=\pi/4$) is new, appears for $\tilde{G}\neq\pm2$, and is described by:
\begin{align}
\hat{\bm{n}}_{1,2}=\frac{1}{\sqrt{2}}\left(i\,,0\,,1\right)\,,\phd\bm{M}(\bm{r})=M\left(\sin(\bm{Q}_1\cdot\bm{r})+\sin(\bm{Q}_2\cdot\bm{r})\,,0\,,\cos(\bm{Q}_1\cdot\bm{r})+\cos(\bm{Q}_2\cdot\bm{r})\right)\ph{\rm and}\ph{\cal F}^{(4)}=2(\beta-\tilde{\beta})\frac{G+\tilde{G}/2}{2}\,.
\end{align}
\subsubsection{C$_4$-symmetric Spin-Whirl Crystal} This double-$\bm{Q}$ C$_4$-symmetric phase is new, appears for $\tilde{G}(1-G)+4G\neq0$, and one finds:
\bea
\hat{\bm{n}}_{1}=\left(i\cos\lambda\,,0\,,\sin\lambda\right)\ph{\rm and}\ph\hat{\bm{n}}_{2}=\left(0\,,i\cos\lambda\,,\sin\lambda\right)\,,\phd 
{\cal F}^{(4)}=2(\beta-\tilde{\beta})\frac{G(\tilde{G}+4)+\tilde{G}}{2(\tilde{G}+4)}\ph{\rm and}\ph\no\\
\bm{M}(\bm{r})=\sqrt{2}M\left(\cos\lambda\sin(\bm{Q}_1\cdot\bm{r})\,,
\cos\lambda\sin(\bm{Q}_2\cdot\bm{r})\,,\sin\lambda\cos(\bm{Q}_1\cdot\bm{r})+\sin\lambda\cos(\bm{Q}_2\cdot\bm{r})\right)\,,\phd\label{eq:MW4a}
\eea

\noi with $\cos(2\lambda)=\tilde{G}/(\tilde{G}+4)$. The latter implies $\cos(2\eta)=0\Rightarrow\eta=\pi/4$. Note also that for $\tilde{g}\,,\tilde{G}=0$ we obtain $\cos(2\lambda)=0\Rightarrow\lambda=\pi/4$ and leads to a symmetric double-$\bm{Q}$ non-coplanar $C_4$-phase $\bm{M}(\bm{r})=M\left(\sin(\bm{Q}_1\cdot\bm{r})\,,\sin(\bm{Q}_2\cdot\bm{r})\,,\cos(\bm{Q}_1\cdot\bm{r})+\cos(\bm{Q}_2\cdot\bm{r})\right)$.

\subsubsection{C$_2$-symmetric Spin-Whirl Crystal} This new double-$\bm{Q}$ C$_2$-symmetric phase, consists of a isotropic magnetic helix for $\bm{Q}_1$ coexisting with an  anisotropic magnetic helix for $\bm{Q}_2$: 
\bea
\hat{\bm{n}}_{1}=\frac{1}{\sqrt{2}}\left(i\,,0\,,1\right)\ph{\rm and}\ph\hat{\bm{n}}_{2}=\left(0\,,i\sin\lambda\,,\cos\lambda\right)\,,\phd
{\cal F}^{(4)}=2(\beta-\tilde{\beta})\frac{(\tilde{G}+4G)^2}{(\tilde{G}+4)^2+16(2G-1)}\qquad\qquad\no\\
\phd{\rm and}\phd\bm{M}(\bm{r})=2M\left(\frac{\cos\eta}{\sqrt{2}}\sin(\bm{Q}_1\cdot\bm{r})\,,\sin\eta\sin\lambda\sin(\bm{Q}_2\cdot\bm{r})
\,,\frac{\cos\eta}{\sqrt{2}}\cos(\bm{Q}_1\cdot\bm{r})+\sin\eta\cos\lambda\cos(\bm{Q}_2\cdot\bm{r})\right)\,,\phd{\rm with}\quad\qquad\\
\cos(2\eta)=-\frac{\tilde{G}^2}{(\tilde{G}+4)^2+16(2G-1)}\quad{\rm and}\quad\cos(2\lambda)=-\tilde{G}\frac{\tilde{G}+4G}{\tilde{G}(\tilde{G}+4)+16G}\,.\qquad\qquad\qquad\qquad
\eea

\section{Induced Charge Order}\label{App:AppendixInducedCharge}

We will here derive the coupling term between magnetic and charge order, that is inducing the latter when magnetism sets in. For simplicity, and without loss of generality, let us restrict ourselves near the magnetic critical temperature, for which an expansion in terms of the magnetic order parameters and the charge density is permissible. We will further consider that the orbital weight, $\hat{v}_{\bm{q}}$, of the magnetic order parameters, $\widehat{\bm{M}}_{\bm{q}}$, is fixed by the spin susceptibility, i.e. $\widehat{\bm{M}}_{\bm{q}}=\bm{M}_{\bm{q}}\hat{v}_{\bm{q}}$ (see Sec.~\ref{Sec:FeSCs}). 

We start from a general multi-orbital model Hamiltonian for an itinerant system, rewritten in $\bm{k}$-space and within the framework of second quantization: ${\cal H}_0=\sum_{\bm{k}}\bm{\psi}_{\bm{k}}^{\dag}\hat{\bm{1}}_2\otimes\widehat{{\cal H}}_0(\bm{k})\bm{\psi}_{\bm{k}}$, where we introduced the electron creation and annihilation operators acting in orbital and spin spaces. Note that $\widehat{{\cal H}}_0(\bm{k})$ is a matrix in orbital space, and the complete Hamiltonian is diagonal in spin space, since SOC has been omitted from the outset. We then add the term: 
\bea
{\cal V}=
\sum_{\bm{k},\bm{q}}\bm{\psi}_{\bm{k}+\bm{q}}^{\dag}\left(\bm{M}_{\bm{q}}\cdot\bm{\sigma}\otimes\hat{v}_{\bm{q}}+\rho_{\bm{q}}\hat{\bm{1}}_2\otimes\hat{\bm{1}}_5\right)
\bm{\psi}_{\bm{k}},
\eea

\noi and perform a perturbative expansion of the resulting free energy to the lowest allowed coupling between charge and magnetic order terms. The lowest order coupling term appears at cubic order of perturbation and the corresponding free energy contribution is
\bea
F=\frac{1}{3\beta}\sum_{\bm{q},\bm{p}}\sum_{ik_n,\bm{k}}
{\rm Tr}\left[\hat{\bm{1}}_2\otimes\widehat{\cal G}_0(ik_n,\bm{k})\widehat{{\cal V}}_{\bm{q}}\hat{\bm{1}}_2\otimes\widehat{\cal G}_0(ik_n,\bm{k}-\bm{q})
\widehat{{\cal V}}_{\bm{p}}\hat{\bm{1}}_2\otimes\widehat{\cal G}_0(ik_n,\bm{k}-\bm{q}-\bm{p})\widehat{{\cal V}}_{-\bm{q}-\bm{p}}\right]\,,\label{eq:QubicPerturbation}
\eea

\noi with the trace being over spin and orbital indices, $1/\beta=k_BT$, $k_n$ denotes the fermionic Matsubara frequencies, $\widehat{\cal G}_0(ik_n,\bm{k})=\big[ik_n\hat{\bm{1}}_5-\widehat{{\cal H}}_0(\bm{k})\big]^{-1}$ defines the bare Matsubara orbital space matrix Green's function and $\widehat{{\cal V}}_{\bm{q}}\equiv\bm{M}_{\bm{q}}\cdot\bm{\sigma}\otimes\hat{v}_{\bm{q}}+\rho_{\bm{q}}\hat{\bm{1}}_2\otimes\hat{\bm{1}}_5$. By carrying out the trace over the spin degrees of freedom and after taking into account all the products of the interaction terms, we have the general result:
\bea
F=\sum_{\bm{q},\bm{p}=\pm\bm{Q}_{1,2}}\left\{\frac{2}{\beta}\sum_{ik_n,\bm{k}}
{\rm Tr}_{\rm orbital}\left[\widehat{\cal G}_0(ik_n,\bm{k})\hat{v}_{\bm{q}}\widehat{\cal G}_0(ik_n,\bm{k}-\bm{q})\hat{v}_{\bm{p}}
\widehat{\cal G}_0(ik_n,\bm{k}-\bm{q}-\bm{p})\right]\right\}|\bm{M}_{\bm{q}}||\bm{M}_{\bm{p}}|\hat{\bm{n}}_{\bm{q}}\cdot\hat{\bm{n}}_{\bm{p}}\ph\rho_{-(\bm{q}+\bm{p})}\,.
\quad
\eea

\noi Therefore we can obtain induced charge order for the set of wavevectors $\{\pm2\bm{Q}_{1,2},\pm\left(\bm{Q}_{1}\pm\bm{Q}_2\right)$, depending on which spin-vector inner products become non-zero. Note that there exist additional higher order contributions to the free energy that yield i) corrections to the above and ii) charge order of additional wavevectors. However, these two contributions will lead to Bragg peaks with lower intensity compared to the one found at cubic order, and thus can be experimentally differentiated.

\section{Magnetoelectric Effects}\label{App:AppendixMagnetoelectricity}

The aim of this paragraph is to illustrate the emergence of the magnetoelectric effects discussed in the main text, whereas the $\bm{q}=\bm{0}$ ferromagnetic component of the magnetization $\bm{M}_{\bm{0}}$, couples to a generally time-dependent electromagnetic vector potential $\bm{A}$, generating an electric field ($\bm{{\cal E}}$) or current ($\bm{I}$). The coupling of the itinerant system to the vector potential will be restricted to the paramagnetic contribution, sufficient for the lowest order coupling with respect to $\bm{A}$ that is pursued here. Therefore, we employ a similar approach as in the paragraph above and focus near the magnetic critical temperature, where the magnetic order parameters can be treated perturbatively. Then we add to the Hamiltonian, in the Schr\"odinger picture, the perturbation term 
\bea
{\cal V}=
\sum_{\bm{k},\bm{q}}\bm{\psi}_{\bm{k}+\bm{q}}^{\dag}\left\{\bm{M}_{\bm{q}}\cdot\bm{\sigma}\otimes\hat{v}_{\bm{q}}+\left[\bm{M}_{\bm{0}}(t)\cdot\bm{\sigma}\otimes\hat{\bm{1}}_5+
\frac{e}{\hbar}\hat{\bm{1}}_2\otimes\frac{\partial\widehat{{\cal H}}_0(\bm{k})}{\partial\bm{k}}\cdot\bm{A}(t)\right]\delta_{\bm{q},\bm{0}}\right\}\bm{\psi}_{\bm{k}}\,,
\eea

\noi with $e>0$. We thus perform a perturbative expansion of the resulting free energy to the lowest allowed coupling between the vector potential, the ferromagnetic moment and the $\bm{q}=\pm\bm{Q}_{1,2}$ magnetic order terms. The lowest order coupling term appears at quartic order of the perturbation term and the corresponding free energy contribution reads
\bea
F&=&\frac{1}{4\beta}\sum_{k,q_{1,2,3}}
{\rm Tr}\left[\hat{\bm{1}}_2\otimes\widehat{\cal G}_0(k)\widehat{{\cal V}}_{q_1}\hat{\bm{1}}_2\otimes\widehat{\cal G}_0(k-q_1)\widehat{{\cal V}}_{q_2}
\hat{\bm{1}}_2\otimes\widehat{\cal G}_0(k-q_1-q_2)\widehat{{\cal V}}_{q_3}
\hat{\bm{1}}_2\otimes\widehat{\cal G}_0(k-q_1-q_2-q_3)\widehat{{\cal V}}_{-q_1-q_2-q_3}\right],\,\qquad\label{eq:QuarticPerturbation}
\eea

\noi where we introduced the four vectors $k=(ik_n,\bm{k})$ and $q=(i\omega_s,\bm{q})$, where the latter consists of the bosonic Matsubara frequency, $i\omega_s$, and wavevector $\bm{q}$. The respective perturbation terms reads 
\bea
\widehat{{\cal V}}_{q}\equiv\bm{M}_{\bm{q}}\cdot\bm{\sigma}\otimes\hat{v}_{\bm{q}}\delta_{\omega_s,0}+\left[\bm{M}_{\bm{0}}(i\omega_s)\cdot\bm{\sigma}\otimes\hat{\bm{1}}_5+
\frac{e}{\hbar}\hat{\bm{1}}_2\otimes\frac{\partial\widehat{{\cal H}}_0(\bm{k})}{\partial\bm{k}}\cdot\bm{A}(i\omega_s)\right]\delta_{\bm{q},\bm{0}}\,.
\eea

\noi For the itinerant systems of interest, inversion symmetry is present, therefore $\widehat{{\cal H}}_0(\bm{k})=\widehat{{\cal H}}_0(-\bm{k})$ and $\partial\widehat{{\cal H}}_0(\bm{k})/\partial k_a=-\partial\widehat{{\cal H}}_0(-\bm{k})/\partial k_a$. As a result, we find that the magnetoelectric coupling can be non-zero only when $\bm{A}||\hat{\bm{q}}$, with $\hat{\bm{q}}=\bm{q}/|\bm{q}|$. By tracing over the spin indices and after taking into account all the possible combinations we obtain:
\bea
F&=&\frac{e}{\hbar}\sum_{\omega_s}\sum_{\bm{q}=\pm\bm{Q}_{1,2}}|\bm{M}_{\bm{q}}|^2
\bm{M}_{\bm{0}}(-i\omega_s)\cdot\left(i\hat{\bm{n}}_{\bm{q}}\times\hat{\bm{n}}_{\bm{q}}^*\right)\hat{\bm{q}}\cdot\bm{A}(i\omega_s)
\frac{2}{\beta}\sum_{k_n,\bm{k}}\no\\
&&\left\{
{\rm Tr}_{\rm orbital}\left[\widehat{\cal G}_0(ik_n,\bm{k})\hat{\bm{q}}\cdot\frac{\partial\widehat{{\cal H}}_0(\bm{k})}{\partial\bm{k}}\widehat{\cal G}_0(ik_n-i\omega_s,\bm{k})
\widehat{\cal G}_0(ik_n,\bm{k})\hat{v}_{\bm{q}}\widehat{\cal G}_0(ik_n,\bm{k}-\bm{q})\hat{v}_{-\bm{q}}\right]\right.\no\\
&&-{\rm Tr}_{\rm orbital}\left[\widehat{\cal G}_0(ik_n,\bm{k})\hat{\bm{q}}\cdot\frac{\partial\widehat{{\cal H}}_0(\bm{k})}{\partial\bm{k}}\widehat{\cal G}_0(ik_n-i\omega_s,\bm{k})\hat{v}_{\bm{q}}\widehat{\cal G}_0(ik_n-i\omega_s,\bm{k}-\bm{q})\widehat{\cal G}_0(ik_n,\bm{k}-\bm{q})\hat{v}_{-\bm{q}}\right]\no\\
&&+\left.
{\rm Tr}_{\rm orbital}\left[\widehat{\cal G}_0(ik_n,\bm{k})\hat{\bm{q}}\cdot\frac{\partial\widehat{{\cal H}}_0(\bm{k})}{\partial\bm{k}}\widehat{\cal G}_0(ik_n-i\omega_s,\bm{k})\hat{v}_{\bm{q}}\widehat{\cal G}_0(ik_n-i\omega_s,\bm{k}-\bm{q})\hat{v}_{-\bm{q}}\widehat{\cal G}_0(ik_n-i\omega_s,\bm{k})\right]\right\}\,.
\eea

\noi In order to investigate the arising magnetoelectric response, one has to perform the analytical continuation, $i\omega_s\rightarrow\omega+i0^+$, to the real frequencies, $\omega$.

\section{Hubbard-Hund model, magnetic order parameters and orbital structure for the case of iron-based superconductors}\label{App:AppendixSusceptibility}

The magnetic transition temperature is obtained by identifying the first zero eigenvalue of the static part of the inverse magnetic propagator matrix when evaluated as a function of temperature and $\bm{q}$. The latter is defined as $\check{{\cal D}}_{\rm mag}^{-1}(\bm{q})= \check{U}^{-1}-\check{\chi}_0(\bm{q})$, with $\check{}$ denoting rank-4 tensors in orbital space. Note that for our calculations we neglect the Hartree shift of the chemical potential induced by interactions. The orbital weight of the magnetic order parameter is obtained from the eigenmatrix associated with the zero eigenvalue, i.e. $\check{{\cal D}}_{\rm mag}^{-1}(\bm{Q}_{1,2})\hat{v}_{1,2}=\hat{0}$. As incommensurability sets in the orbital content acquires imaginary parts, e.g. for LaFeAsO we find:
\bea
	\hat{v}_{1} = \begin{pmatrix}
		0.395	& 		0 	& 	i0.045 	& 	0	 	&	 0 \\
		0 		&	0.501 	& 	0 		& 	i0.028 	& 	-i0.025 \\
		i0.045  & 	0 		& 	0.574   &  	0 		&  		0 \\
		0 		& 	i0.028   & 	0 	 	&  	0.335   &  	-0.038 \\
		0  		& 	-i0.025 & 	0 		& 	-0.038 	& 	0.377
	\end{pmatrix}\quad{\rm and}\quad
	\hat{v}_{2} = \begin{pmatrix}
		0.501	& 		0 	& 	0 	& 	i0.028	 	&	 i0.025 \\
		0 		&	0.395 	& 	-i0.045 		& 	0 	& 	  0 \\
		0  & 	-i0.045 		& 	0.574   &  	0 		&  		0 \\
		i0.028 		& 	0   & 	0 	 	&  	0.335   &  	0.038 \\
		i0.025 		& 	0 & 	0 		& 	0.038 	& 	0.377
	\end{pmatrix}\,,\quad\label{eq:Mag}
\eea

\noi where $\hat{v}_1$ is plotted in Fig.~\ref{fig:Susc_FS_Orb_W}(l) in the main text. The quartic coefficients of the free energy are computed by performing a Hubbard-Stratonovich decoupling in the magnetic channel and expanding the trace-log to fourth order in the magnetic order parameters. The expression is truncated by assuming that only the lowest harmonics contribute. This is justified by comparing the magnitude of the peak at $\bm{Q}_{1,2}$ in the bare susceptibility with the peaks at higher integer multiples of $\bm{Q}_{1,2}$, as shown in Figs.~\ref{fig:Susc_FS_Orb_W}(d) and (j). The quartic coefficients are rank-8 tensors in orbital space and to determine the magnetic order at the instability the coefficients are projected onto the leading instability using the orbital content provided by the above eigenmatrices.

The expressions for the quartic coefficients of the free energy are related to the microscropic bandstructure through
\bea
	\tilde{\beta}_1^{abcdefgh} &=& \frac{1}{16\beta{\cal N}} \sum_{k} \big[ G^{ab}G^{cd}_1 G^{ef} G^{gh}_1 + G^{eb}G^{ch}_1 G^{af} G^{gd}_1 + 2 G^{gb} G^{cd}_1 G^{eh} G^{af}_{-1} - 2 G^{gb}G_{1}^{ch}G^{ad}G^{ef}_{-1} \big]\label{eq:betatilde1} \\
	\tilde{\beta}_2^{abcdefgh} &=& \frac{1}{16\beta{\cal N}}\sum_{k}\big[ G^{ab}G^{cd}_2 G^{ef} G^{gh}_2 + G^{eb}G^{ch}_2 G^{af} G^{gd}_2 + 2 G^{gb} G^{cd}_2 G^{eh} G^{af}_{-2} - 2 G^{gb}G_{2}^{ch}G^{ad}G^{ef}_{-2} \big] \\
	\left(\beta_1 - \tilde{\beta}_1 \right)^{abcdefgh} &=& \frac{1}{16\beta{\cal N}}\sum_{k}\big[ - G^{ab}G^{cf}_1 G^{gd} G^{eh}_1 + 2 G^{eb}G^{cf}_1 G^{gh} G^{ad}_{-1} \big] \\
	\left( \beta_2 - \tilde{\beta}_2 \right)^{abcdefgh} &=& \frac{1}{16\beta{\cal N}}\sum_{k}\big[ - G^{ab}G^{cf}_2 G^{gd} G^{eh}_2 + 2 G^{eb}G^{cf}_2 G^{gh} G^{ad}_{-2} \big] \\
	g^{abcdefgh} &=& \frac{1}{8\beta{\cal N}}\sum_{k}\big[ G^{ab}G^{cd}_1 G^{ef} G^{gh}_2 + G^{gb}G^{cd}_1 G^{eh} G^{af}_{-2} + G^{ad}G^{eb}_{-1}G^{cf}G^{gh}_{2} + G^{gd}G^{eb}_{-1}G^{ch}G^{af}_{-2} \nonumber \\
	&& \qquad \qquad - G^{ef}G^{gb}_{2}G^{ch}_{1+2}G^{ad}_1 - G^{ch}G^{ad}_{1}G^{ef}_{1+2}G^{gb}_{2} \big] \\
	\tilde{g}_1^{abcdefgh} &=& \frac{1}{8\beta{\cal N}}\sum_{k}\big[ -G^{ab}G^{cf}_1 G^{gd} G^{eh}_2 + G^{eb}G^{cf}_1 G^{gh} G^{ad}_{-2} + G^{af}G^{gb}_{-1}G^{cd}G^{eh}_{2} - G^{ef}G^{gb}_{-1}G^{ch}G^{ad}_{-2} \nonumber \\
	&& \qquad \qquad + G^{gd}G^{eb}_{2}G^{ch}_{1+2}G^{af}_1 + G^{ab}G^{cd}_{1}G^{ef}_{1+2}G^{gh}_{2} \big] \\ \nonumber
	\tilde{g}_2^{abcdefgh} &=& \frac{1}{8\beta{\cal N}}\sum_{k}\big[ G^{eb}G^{cf}_1 G^{gh} G^{ad}_2 - G^{ab}G^{cf}_1 G^{gd} G^{eh}_{-2} - G^{ef}G^{gb}_{-1}G^{ch}G^{ad}_{2} + G^{af}G^{gb}_{-1}G^{cd}G^{eh}_{-2} \nonumber \\
	&& \qquad \qquad + G^{gh}G^{ab}_{2}G^{cd}_{1+2}G^{ef}_1 + G^{eb}G^{ch}_{1}G^{af}_{1+2}G^{gd}_{2} \big]\,,\label{eq:gtilde2}
\eea

\noi with ${\cal N}$ denoting the number of sites in $\bm{k}$-space. Here $k=(ik_n,\bm{k})$ and $G^{ab}_{1,2}$ is the fermionic Green function centered at $\bm{Q}_{1,2}$:
\begin{eqnarray}
	G^{ab}_{1,2} &=& \sum_{\nu} \frac{u^{a}_{\nu}(\bm{k}+\bm{Q}_{1,2}) u^{b}_{\nu}(\bm{k}+\bm{Q}_{1,2})^{\ast}}{ik_n - E_{\nu}(\bm{k}+\bm{Q}_{1,2})}\,,
\end{eqnarray}
with $k_n$ a fermionic Matsubara frequency. In the above we also employed the shorthand notation $f_{1+2}\equiv f_{\bm{Q}_1+\bm{Q}_2}$.

Note that the tensorial coefficients, e.g. $\tilde{\beta}_1^{abcdefgh}$ and $\tilde{\beta}_2^{abcdefgh}$, are not identical but are related by $C_4$ rotations. To proceed we project these onto the leading magnetic instability utilizing the orbital content provided by the matrices $v_{ab}(\bm{Q}_{1,2})$ as in Ref.~\onlinecite{christensen17}, for instance we have
\bea
\beta_1 - \tilde{\beta}_1 \equiv \sum_{\substack{abcd \\ efgh}}(\beta_1 -\tilde{\beta}_1)^{abcdefgh} v_{ab}(\bm{Q}_1)v_{cd}(\bm{Q}_{1})v^{\ast}_{ef}(\bm{Q}_1)v^{\ast}_{gh}(\bm{Q}_{1})\,.
\eea
Note that the contracted coefficients, such as $\beta_1 - \tilde{\beta}_1$ and $\beta_2 - \tilde{\beta}_2$ are identical. 

\end{widetext}


\begin{thebibliography}{}
\bibitem{SkyrmionComment} $\Theta$ is a sym\-metry of the IC magnetization profile alone since, due to the IC nature of $\bm{Q}_{1,2}$, the involved translations are not contained in the ones given by the crystalline periodicity of the FeSCs lattice. Ne\-ver\-theless, as long as the low-energy properties of these itinerant magnets are primarily dictated by the Fermi-surface electrons, we expect $\Theta$ to be effectively present. This is because the electrons near the Fermi-surface can be described by an effective continuum model which is unaware of the lattice constant.
\bibitem{CommensurateLimit} Notably, taking the commensurate limit for the magnetic profiles of DPHM and SWC$_4$, without any further considerations, is ambiguous since \{DPMH, SWC$_4$\}$\rightarrow$\{CSDW, SVC\}. This is due to the translational invariance of the system and our freedom to pick alternative spatial profiles other than the ones chosen in Sec.~\ref{Sec:ICMagneticPhases}. Nonetheless, by additionally demanding that the free-energy minimum varies continuously when taking this limit, Fig.~\ref{fig:GenPhaseDiagram}(b) yields DPMH$\rightarrow$CSDW and SWC$_4\rightarrow$SVC.
\bibitem{NoteSOC1} For a weak SOC we can approach the problem u\-sing the already-obtained set of magnetic-order-parameter extrema of Eq.~\eqref{eq:free_energy}. However, if the SOC becomes strong, one should minimize the combination of Eqs.~\eqref{eq:free_energy} and~\eqref{eq:free_energy_SOC}, that will generally lead to a new set of extrema.
\bibitem{NoteSOC2} Note that the potential inclusion of SOC terms at quartic order would lower $T_{\rm mag}^{(2)}$ and if they were sufficiently strong, they could even prevent the occurence of the se\-con\-da\-ry transition. However, such as situation is beyond the scope and applicability of the present analysis.
\end{thebibliography}
\end{document}